\documentclass[pra,twocolumn,showpacs,amsmath,amssymb,amsfonts]{revtex4}
\begin{document}

\title{Canonical averaging of the equations of quantum mechanics}

\author{A.G.Chirkov}

\email{agc@AC11593.spb.edu}

\affiliation{Department of Theoretical Physics, State Polytechnic
University, St.-Petersburg, 195251, Russia}

\date{\today}

\begin{abstract}

The representation of a Schrodinger equations as a classic
Hamiltonian system allows to construct a unified perturbation
theory both in classic, and in a quantum mechanics grounded on the
theory of canonical transformations, and also to receive
asymptotic estimations of affinity of the precisian approximated
solutions of Schrodinger equations
\end{abstract}


\maketitle


\section{Introductory remarks}

The very possibility of applying the modern methods of the classical theory
of non-linear oscillations to quantum mechanics is based upon the
representation of the non-stationary Schr\"{o}dinger's equation as a
\index{Schr\"{o}dinger's equation}classical Hamiltonian system. From this
perspective it is quite natural to construct a special asymptotic
perturbation theory which utilises the advantages of the Hamiltonian
formalism, that is to apply canonical transformations. Special and
sufficiently efficient approaches \cite{30}, \cite{602} and \cite{603} were
developed by mathematicians for canonical systems. However, the generality
of these approaches makes them very cumbersome whereas the first two
non-trivial approximations are ordinarily sufficient for practical
application.

Traditionally, mathematicians use the methods of spectral analysis of
operators for constructing perturbation theory in non-relativistic quantum
mechanics, see \cite{604}, \cite{605}. However, it is necessary to take into
account that, in practice, physicists do not distinguish between the
concepts of self-adjoint and symmetric operators. This gives rise to two
unpleasant things: firstly, the domain of definition of the operator in
Hilbert space remains unclear which does not allow one to apply the methods
of spectral theory, and secondly, the domain of definition of the operator
includes all functions for which analytical operations are meaningful
regardless of the fact whether these functions (and the result of applying
an operator to them) belong to%
\index{Hilbert space} Hilbert space. A rigorous consideration of the latter
case requires the introduction of an equipped Hilbert space, see \cite{606},
\cite{607}.

For this reason it is not possible to prove even the conditions for
applicability of the regular perturbation theory developed by Kato-Relih
\cite{605}, \cite{608} providing us with the criterion for the
Rayleigh-Schr\"{o}dinger formal series to have a non-zero radius of
convergence.

It is well known that orthodox perturbation theory is not applicable to many
cases since the corresponding series diverge. The asymptotic character of
the series used in perturbation theory was first proved by Titchmarsh \cite
{609}. A rigorous proof of the divergence of this series for an anharmonic
oscillator $(V\sim x^{4})$ is given by Bender and Woo \cite{610}.

In addition to the associated complexity, the above-mentioned methods
possess another shortcoming, namely they do not allow one to obtain the wave
function which plays an important role in the investigation of physical
systems.

In the present chapter we suggest another approach which is based upon the
representation of the non-stationary Schr\"{o}dinger's equation as a
classical Hamiltonian system. This representation enables one to make use of
the powerful, modern, rigorously substantiated methods of the classical
theory of non-linear oscillations (asymptotic perturbation theory) and
indicate simple conditions for justifying the applicability of the results
obtained.

An important part for the transition from the hypotheses of Planck and
Einstein to quantum mechanics was played by the adiabatic hypothesis by
Ehrenfest. Born and Fock showed in 1928, \cite{611}, that Ehrenfest's
hypothesis is a consequence of the postulates of quantum mechanics. A
rigorous mathematical proof of the adiabatic theory was given by Kato in
1949, \cite{612}. Later on, the adiabatic Landau-Dykhne approximation, \cite
{613}, \cite{614}, \cite{615}, was built on the analogy between the
adiabatic and quasi-classical approximations.

The Born-Fock adiabatic approximation
\index{adiabatic approximation}is actually not an approximation since all of
the terms of the adiabatic Born-Fock series have the same order of
smallness, \cite{615}, \cite{616}, which, in turn, does not allow us to
construct a post-adiabatic approximation. The Born-Fock condition, which
implies real-valued wave functions, does not allow us to use this
approximation in problems involving magnetic fields.

The results of the Landau-Dykhne adiabatic approximation relate to the
results of non-stationary perturbation theory only approximately. In
addition to this, both approximations yield an incorrect factor in front of
the exponential function, see \cite{616}.

The problem of the time interval, within which the difference between the
approximated and exact solutions is small, plays an important part in the
non-stationary case. In the above works, this problem is not discussed at
all.

In the present chapter, the adiabatic and the post-adiabatic theories, as
well as the adiabatic perturbation theory and the post-adiabatic
approximation, are constructed by means of the method of canonical averaging
(phase perturbation theory). The basic assumptions of the Born-Fock theory
are not satisfied. The obtained approximations are compared with exact
solutions of the non-stationary Schr\"{o}dinger's equation for a harmonic
oscillator in a homogeneous time-dependent field and with the approximations
obtained by traditional formulae \cite{613}. One can see from this
comparison that the standard non-stationary approximations \cite{613} are
valid only within a non-dimensional time intervals $t\sim 1$, whereas the
approximations of the present chapter are valid within asymptotically longer
time intervals $t\sim 1/\varepsilon $.

\section{Stationary Schr\"{o}dinger's equation as a classical Hamiltonian
system}

In this section the classical canonical perturbation theory is applied to
constructing asymptotic solutions of
\index{Schr\"{o}dinger's equation!stationary}Schr\"{o}dinger's equation with
a discrete spectrum. The main subject of analysis of the non-relativistic
quantum theory is Schr\"{o}dinger's equation, \cite{613},
\begin{equation}
i\hbar
\frac{\partial \Psi (q,t)}{\partial t}=\hat{H}\Psi (q,t),  \label{eq1}
\end{equation}
where $i^{2}=-1,\quad\hbar =1.054\ 10^{-34}\emph{J}\emph{s}$
denotes Planck's constant, $q=(q_{1},q_{2},$ $...,q_{n})$ denotes
a point of the
configuration space of the corresponding classical system, $t$ is time and $%
\Psi (q,t)$ denotes a complex-valued function with integrable square of the
absolute value. Further, $\hat{H}$ denotes a self-adjoint (symmetric)
operator in
\index{Hilbert space}Hilbert space, which in Cartesian coordinates, in
Schr\"{o}dinger's representation for one particle, has the form, \cite{613},
\begin{equation}
\hat{H}=\hat{T}+\hat{V}=-\frac{{\hbar ^{2}}}{{2}m}\Delta +\hat{V}(x,y,z),
\label{eq2}
\end{equation}
where $\hat{T}$ and $\hat{V}$ designate operators of the kinetic and
potential energies, respectively, $\Delta =\dfrac{\partial ^{2}}{\partial
x^{2}}+\dfrac{\partial ^{2}}{\partial y^{2}}+\dfrac{\partial ^{2}}{\partial
z^{2}}$, and $m$ is mass of the particle. Schr\"{o}dinger's equation (\ref
{eq1}) is subject to an initial condition $\Psi (q,0)=\Psi _{0}(q)$ and some
boundary conditions.

The case studied in the framework of perturbation theory appears when
operator $\hat{H}$ can be cast as the sum
\begin{equation}
\hat{H}=\hat{H}_{0}+\varepsilon \hat{V},\quad 0<\varepsilon <<1,  \label{eq3}
\end{equation}
of two
\index{self-adjoint operator}self-adjoint operators, the corresponding
problem (\ref{eq1}) for operator $%
\hat{H}_{0}$ being assumed to have an exact solution and the second operator
(perturbation) being small in some sense, \cite{605}, \cite{608}, \cite{613}.

The majority of the physically interesting problems turn out to be
mathematically incorrect, since the perturbation operators are usually not
boun{-}ded and not even self-adjoint. The latter is related to the fact that
physicists never distinguish between the concepts of self-adjoint and
symmetric operators. This leads to the operator space in Hilbert space being
unclear, which in turn does not allow one to apply the methods of spectral
theory \cite{607}, \cite{608}. For this reason, it is difficult to indicate
the conditions for the applicability of perturbation theory and estimate the
discrepancy between the exact and an approximate solution for practical
problems.

However, it is possible to reduce Schr\"{o}dinger's equation to a form of
classical Hamiltonian system, which is well-developed in non-linear
mechanics. This allows one to apply the methods of classical dynamics which
are rigorously substantiated and simpler from the perspective of application.

Let us consider eq. (\ref{eq1}) with Schr\"{o}dinger's operators (\ref{eq3}%
), i.e. the problem
\begin{eqnarray}
i\hbar \frac{{\partial \Psi (q,t)}}{{\partial t}} &=&(\hat{H}_0+\varepsilon
\hat{V})\Psi (q,t),  \nonumber \\
\Psi (q,0) &=&\Psi _0(q),  \label{eq4}
\end{eqnarray}
where operator $\hat{H}_0$ does not depend on time and $\varepsilon $ is a
formal small parameter. The question of choosing the small parameters is
discussed below.

Along with the problem we consider, the generating approximation, which is
obtained from eq. (\ref{eq4}) at $\varepsilon =0$%
\begin{equation}
i\hbar \frac{{\partial \Psi ^{0}(q,t)}}{{\partial t}}=\hat{H}_{0}\Psi
^{0}(q,t),\;\;\;\;\;\;\Psi ^{0}(q,0)=\Psi _{0}(q)\,.  \label{eq5}
\end{equation}
Assuming the spectrum to be discrete, we can apply Fourier's method and set
the general solution of problem (\ref{eq5}) in the form
\begin{eqnarray}
\Psi ^{0}(q,t) &=&\sum\limits_{n=0}^{\infty }c_{n}^{0}\psi _{n}^{0}(q)\exp
(-i\omega _{n}^{0}t),  \nonumber \\
c_{n}^{0} &=&\int \Psi _{0}(q)\psi _{n}^{0\ast }(q)dq,\quad \omega
_{n}^{0}=E_{n}^{0}/\hbar ,  \label{eq6}
\end{eqnarray}
where $\psi _{n}^{0}(q)$ and $E_{n}^{0}$ denote respectively the
eigenfunctions and eigenvalues of the following problem
\begin{equation}
\hat{H}_{0}\psi _{n}^{0}(q)=E_{n}^{0}\psi _{n}^{0}(q)  \label{eq7}
\end{equation}
and an asterisk denotes the complex conjugate.

The self-adjoint character of operator $\hat{H}_0$ means that for $\Psi
(q,t)\in L^2$ the following expansion is valid
\begin{equation}
\Psi (q,t)=\sum\limits_{n=0}^\infty c_n(t)\psi _n^0(q)\exp (-i\omega
_n^0t)\,.  \label{eq8}
\end{equation}

Inserting expansion (\ref{eq8}) into eq. (\ref{eq4}) yields the following
equations for the coefficients of the expansion $c_n(t),c_n^{*}(t)$

\begin{eqnarray}
\dot{c}_{n}(t) &=&-i\varepsilon \sum\limits_{m=0}^{\infty
}v_{nm}c_{m}(t)\exp (-i\omega _{mn}^{0}t),  \nonumber \\
\dot{c}_{n}^{\ast }(t) &=&i\varepsilon \sum\limits_{m=0}^{\infty
}v_{mn}c_{m}^{\ast }(t)\exp (i\omega _{mn}^{0}t),  \label{eq9}
\end{eqnarray}
\begin{equation}
v_{mn}(t)=\frac{{1}}{{\hbar }}\int \psi _{m}^{0\ast }(q)V(q,t)\psi
_{n}^{0}(q)dq\,,  \label{eq10}
\end{equation}
where a dot implies time derivative and $\omega _{mn}^{0}=\omega
_{m}^{0}-\omega _{n}^{0}.$

The system of equations (\ref{eq9}) is Hamiltonian (in the classical sense)
with the following Hamilton function
\begin{equation}
\varepsilon H_{1}(c,c^{\ast },t)=-i\varepsilon \sum\limits_{n,m=0}^{\infty
}v_{nm}c_{n}^{\ast }c_{m}\exp (i\omega _{nm}^{0}t)  \label{eq11}
\end{equation}
and describes a classical distributed system with an infinite number of
internal resonances. The system is Hamiltonian as matrix $v_{nm}$ is
Hermitian, that is the perturbation operator is self-adjoint. By separating
the principal resonance $\omega _{n}^{0}=\omega _{m}^{0}$, we can cast
Hamilton's function (\ref{eq11}) as follows
\begin{eqnarray}
\varepsilon H_{1}(c,c^{\ast },t)&=&-i\varepsilon
v_{nn}c_{n}c_{n}^{\ast }
\nonumber\\
& &-i\varepsilon {\sum\limits_{n,m=0}^{\infty }{^{\prime
}}}v_{nm}c_{n}^{\ast }c_{m}\exp (i\omega _{nm}^{0}t)\,
\label{eq12}
\end{eqnarray}

\noindent where a prime denotes the sum without the term with
$n=m$. Transformation of variables $c_{n},c_{n}^{\ast }$ to the
real-valued ''action-angle'' variables $I_{n},\psi _{n}$ by means
of the formulae
\begin{eqnarray}
c_{n} &=&\sqrt{I_{n}}\exp (-i\psi _{n}),\qquad
I_{n}=c_{n}c_{n}^{\ast },
\nonumber \\
c_{n}^{\ast } &=&\sqrt{I_{n}}\exp (i\psi _{n}), \psi _{n}=-\arctan
\frac{{(c_{n}-c_{n}^{\ast })}}{{i(c_{n}+c_{n}^{\ast })}}
\label{eq13}
\end{eqnarray}

\noindent we can set system (\ref{eq9}) in the form

\begin{equation}
\begin{gathered}
\mathop {I_n }\limits^ \bullet   = \varepsilon 2
\sum\limits_{m = 0}^\infty  {'\sqrt {I_n I_m } }  \hfill \\
\,\,\,\,\,\,\,\,\,\,\,\,\,\,\,\,\,\,\,\,\,\,\,\,\,\,\,\,\,
\times \,\,{\text{Im}}\{ v_{nm} {\text{exp}}
[ - i(\psi _m  - \psi _n  + \omega _{mn}^0 t)]\}  \hfill \\
\mathop {\psi _n }\limits^ \bullet   = \varepsilon v_{nn}
+ \varepsilon \sum\limits_{m = 0}^\infty {'\sqrt {{{I_n }\mathord{\left/{\vphantom {{I_n } {I_m }}} \right.
\kern-\nulldelimiterspace} {I_m }}}}  \hfill \\
\,\,\,\,\,\,\,\,\,\,\,\,\,\,\,\,\,\,\,\,\,\,\,\,\,\,\,\,\,
\times \,\,{\text{Re}}\{ v_{nm} {\text{exp}}
[ - i(\psi _m  - \psi _n  + \omega _{mn}^0 t)]\}  \hfill \\
\end{gathered} \label{eq14}
\end{equation}

\noindent and Hamilton's function (\ref{eq12}) as follows

\begin{equation}
\begin{gathered}
\varepsilon H_1 (I,\psi ,t) = \varepsilon \sum\limits_{m,n = 0}^\infty  {v_{nm} \sqrt {I_n I_m } }  \hfill \\
\,\,\,\,\,\,\,\,\,\,\,\,\,\,\,\,\,\,\,\,\,\,\,\,\,\,\,\,\,\,\,\,\,\,\,\,\,\,\,\,\,\,\,\,\,\,\, \times \,\,exp[ - i(\psi _m  - \psi _n  + \omega _{mn}^0 t)] \hfill \\
\end{gathered} \label{eq15}
\end{equation}

Systems (\ref{eq9}) and (\ref{eq14}) do not contain Planck's constant
explicitly and are the classical Hamiltonian systems with an infinite number
of internal resonances. Estimates of the norm of discrepancy between the
exact and approximate solutions as well as the conditions for applicability
of the averaging method for these systems are given by the Los theorem, see
\cite{617} and \cite{618}, which is a generalisation of Bogolyubov theorem
for the case of an infinite-dimensional coordinate Hilbert space.

The canonical form of systems (\ref{eq9}) and (\ref{eq14}) allows us to
consider the evolutionary equations by operating only with Hamilton's
functions (\ref{eq12}) and (\ref{eq15}), i.e. by calculating an averaged
Hamilton's function. For example, for eqs. (\ref{eq11}) and (\ref{eq15}) the
second approximation $\bar{H}^{(2)}$ for the averaged Hamilton's function is
constructed with the help of the following formulae
\begin{gather}
\bar{H}^{(2)}(\bar{c},\bar{c}^{\ast })=\varepsilon \bar{H}_{1}(\bar{c},\bar{c%
}^{\ast })+\varepsilon ^{2}\bar{H}_{2}(\bar{c},\bar{c}^{\ast }),  \nonumber
\\
\bar{H}_{1}=\langle H_{1}\rangle ,\quad \bar{H}_{2}=-\left\langle \frac{%
\partial \tilde{H}_{1}}{{\partial \bar{c}^{\ast }}}\frac{{\partial }\left\{
H_{1}\right\} }{{\partial \bar{c}}}\right\rangle ,  \label{eq16}
\end{gather}
where $\bar{c}_{n}$ and $\bar{c}_{n}^{\ast }$ denote the evolutionary
components of variables $c_{n}$\ and $c_{n}^{\ast }$. In the latter equation
the following notation is used
\begin{eqnarray}
\langle f\rangle &=&\lim_{T\rightarrow \infty }\frac{1}{T}%
\int\limits_{t_{0}}^{t_{0}+T}f(\bar{c},\;\bar{c}^{\ast },t)dt,
\nonumber \\
\tilde{f}%
(\bar{c},\;\bar{c}^{\ast },t)&=&f(\bar{c},\;\bar{c}^{\ast
},t)-\langle
f\rangle ,  \nonumber \\
\left\{ f\right\}&=&\int \tilde{f}(\bar{c},\;\bar{c}^{\ast },t)dt,
\label{eq17}
\end{eqnarray}
the arbitrary function of slow variables $\bar{c}\;$and $\bar{c}^{\ast }$
being set to zero while estimating the last integral.

Let us notice that $\bar{H}$ is an integral of the averaged equations of
motion, i.e. an adiabatic invariant \cite{620}, \cite{621}.

The first approximation $\bar{c}_{n}^{(1)}$ to expansion coefficient $c_{n}$
is given by the formula $c_{n}^{(1)}=\bar{c}_{n}$, where $\bar{c}_{n}$
satisfies the following equation
\begin{equation}
\stackrel{\bullet }{\bar{c}}_{n}=\varepsilon \frac{{\partial }\bar{H}_{1}}{%
\partial \bar{c}_{n}^{\ast }}\;.  \label{eq18}
\end{equation}

The second approximation $c_{n}^{(2)}$ to expansion coefficient $c_{n}$ is
given by
\begin{equation}
\bar{c}_{n}^{(2)}=\bar{c}_{n}+\varepsilon \frac{{\partial {\left\{ {H_{1}}%
\right\} }}}{{\partial \bar{c}_{n}^{\ast }}},  \label{eq19}
\end{equation}
where the second approximation to evolutionary component $\bar{c}_{n}$ is
obtained from the equation
\begin{equation}
\stackrel{\bullet }{\bar{c}}_{n}=\varepsilon \frac{{\partial }\bar{H}_{1}}{{%
\partial }\bar{c}_{n}^{\ast }}+\varepsilon ^{2}\frac{{\partial }\bar{H}_{2}}{%
\partial \bar{c}_{n}^{\ast }}\;.  \label{eq20}
\end{equation}

\section{General properties of the canonical form of Schr\"{o}dinger's
equation}

Representation of Schr\"{o}dinger's equation
\index{Schr\"{o}dinger's equation}in canonical form (\ref{eq9}) or (\ref
{eq14}) allows us to draw a number of conclusions without performing any
calculations.

1. The original formulation of the problem of perturbation theory for the
non-stationary Schr\"{o}dinger's equation results in formulae enabling us to
study all cases: stationary (non-degenerate and degenerate), non-stationary,
resonant, adiabatic etc.

2. In the stationary case, for which matrix elements $v_{mn}$ are
independent of time, in the absence of degeneracy and the internal
resonances ($\omega _{mn}^{0}\neq O\left( {\varepsilon }\right) ,\;m\neq n$)
except for the selected principal resonance $(m=n)$, systems (\ref{eq9}) or (%
\ref{eq14}) admit direct averaging. The equation for the evolutionary
components of phase\ $%
\bar{\psi}_{n}$ has the form $\dot{\psi}_{n}=\varepsilon v_{nn}$ (the mean
values of the sums in eqs. (\ref{eq9}) or (\ref{eq14}) are zero) and is
easily integrated, to give $\bar{\psi}_{n}=\varepsilon v_{nn}t+\psi _{n0}$.
In turn, this means that the first correction to the energy appears due to
the principal internal resonance. This conclusion remains valid for any
order of the perturbation theory since the corrections to the eigenvalues of
the unperturbed operator $\hat{H}_{0}$ are basically determined by the
principal internal resonance.

On the other hand, it is evident that the coefficients of the diagonal terms
in the averaged Hamilton function are the corrections to the energy.

Let us notice that direct averaging without introducing the resonant terms
would lead to the appearance of divergent terms which are proportional to $%
1/\left( E_m-E_n\right) $ $(m\rightarrow n)$. Thus, separation of the
resonant terms eliminates the divergent terms in the perturbation theory
series.

3. It is obvious that the first-order contribution to the averaged Hamilton
function or to the above equations from the non-resonant sums is zero for a
stationary perturbation and non-zero for a non-stationary perturbation. This
is the only difference between the stationary and non-stationary cases. In
both cases, calculations are carried out by means of formula (\ref{eq16}).

4. Contributions from these sums are not zero if the perturbation is
stationary and the unperturbed levels contain close levels for which $\omega
_{mn}^{0}=O\left( {\varepsilon }\right) $. Thus, the problem of close levels
should be solved only in a non-stationary form. It is evident that the
problem for degenerate levels is a particular case of the previous one for
which, along with the principal quantum number $n$, there is a multiindex $%
\alpha $ characterising the unperturbed eigenvalue, so that the relationship
$\omega _{n\alpha ,n\beta }=O\left( {\varepsilon }\right) $ holds. All these
cases are manifestations of the resonances additional to the principal
internal resonance.

5. Generally speaking, a general analysis is not applicable for
time-dependent perturbations because it is necessary to know the spectrum of
the perturbation so as to take correct account of the possible resonances.
Only general analysis of periodic (in particular, single-frequency)
perturbation is feasible.

6. For the sake of simplicity, let us consider the case of single-frequency
perturbation ($v_{mn}\sim \exp (\pm i\omega t))$. Clearly, the problem in
this case is reduced to the stationary one, for which the expansion is
performed for wave functions of the stationary states with the new
frequencies $\omega _{mn}=\omega _{m}-\omega _{n}$ such that $\omega
_{m}=\omega _{m}^{0}-\omega /2,\;\omega _{n}=\omega _{n}^{0}+\omega /2$ $($%
for $v\sim \exp (i\omega t))$ and $\omega _{m}=\omega _{m}^{0}+\omega
/2,\;\omega _{n}=\omega _{n}^{0}-\omega /2$ $($ for $v\sim \exp (-i\omega
t)) $.

Thus, the corresponding quantities $E_{m,n}=\hbar \omega _{m,n}$ are
energies, that is the energy of the system: unperturbed system (atom) and a\
field. Their interaction is absent, the original interaction being included
into the definition of the quasi-energy. This gives rise to the concept of
the system dressed by a field (dressed atom, \cite{621}).

Traditionally, this conclusion is obtained in a rather sophisticated way by
means of Floquet theorem and the conclusion on the level splitting is
obtained only in the resonant approximation, see \cite{615}, \cite{620}. Let
us notice that the results of this point are valid for any value of
parameter $\varepsilon $, that is regardless of the perturbation theory.

7. The exact eigenfrequencies of the perturbed system are given by the
relationship $\omega _{n}=\omega _{n}^{0}+\dot{\psi}_{n}$. Thus, the second
equations in (\ref{eq14}) determine corrections to the eigenfrequencies
caused by perturbation. In principle, these corrections can be removed by a
canonical transformation of the phase which can serve as a distinctive
procedure of renormalisation which allows one to remove the secular terms
from the series of the perturbation theory. It is obvious that in the
non-stationary case (even for $v_{nn}=0)$ there exists a non-zero
contribution of the first order stemming from the sum. The presence of this
contribution is not possible using orthodox perturbation theory. The
coherent interaction with the external field is realised under the condition
of constant phase difference (the condition of equality of the original
frequencies is only a necessary condition) and has the form $\dot{\varphi}%
_{2}-\dot{\varphi}_{1}-\omega =\omega _{2}^{0}+\dot{\psi}_{2}-\omega
_{1}^{0}+\dot{\psi}_{1}=\omega _{21}^{0}-\omega +\dot{\psi}_{2}-\dot{\psi}%
_{1}=0$, where $\omega $ denotes the frequency of the external field
satisfying the condition that $\omega _{21}^{0}-\omega =O\left( {\varepsilon
}\right) $.

One usually uses the condition that the transition frequency is close to
that of the external field. It follows from the form of the equations for $%
\dot{\psi}_{n}$ that the type of problem for resonant interaction coincides
with the type considered in point 4 above.

The exact frequencies $\omega _n $n the system are anisochronous which is a
characteristic of the non-linear classical system and leads to bounded
solutions at resonance even in the case of no damping, in spite of the
linearity of  Schr\"{o}dinger's equation.

Thus, Schr\"{o}dinger's equation is equivalent to some classical non-linear
distributed system whereas representations (\ref{eq9}) or (\ref{eq14}) are
expansions in terms of normal forms of the unperturbed system.

To some extent, it is the picture to which Schr\"{o}dinger tended
and which is most close to the classical one. ''There is no need
to explain that the representation of the energy transformation
from one oscillatory form to another under a quantum energy
transition is much more satisfactory than that of electron
jump''$^1$\footnotetext [1]{Translation from Russian}, \cite{623}.

8. It is evident that no specific ''quantum-mechanical'' properties of the
generating operator $\hat{H}_0$, but being self-adjoint, is used.
Nevertheless, this technique allows us to analyse other self-adjoint
problems of mathematical physics with a discrete spectrum. Taking into
account the particular structure of the generating operator $\hat{H}_0$ we
can construct a more efficient theory \cite{624}.

An attempt to apply the averaging method to quantum mechanics was undertaken
in \cite{625}. However, as follows from the above, the absence of resonances
(the main assumption of the authors) is not correct. Besides, despite the
title of this article, the authors did not succeed in a proof of the theorem
of convergence of the constructed perturbation theory.

9. All formulae remain valid for the case of adiabatic perturbation, i.e.
under the additional dependence of perturbation on the slow time $\tau
=\varepsilon t$ $(\varepsilon V=\varepsilon V\left( {\vec{r},t,\tau }\right)
)$. In this case, equations for the evolutionary components become
non-stationary and require more sophisticated integration methods.

10. Small parameter $\varepsilon $ is introduced in systems
(\ref{eq9}) and (\ref{eq14}) in a formal way. Generally speaking,
the question of a rigorous introduction into equations should be
considered individually for each particular problem. Let us point
out some general ideas.

Let us introduce some characteristic values $\left[ {E}\right] $ and $\left[
{V}\right] $ for the eigenvalues $E$ and matrix elements $V_{mn}$
respectively. Then $\left[ {E}\right] /\hbar =\omega _0$ and $\left[ {V}%
\right] /\hbar =\Omega _0$ can be referred to as a characteristic
eigenfrequency and the generalised Rabi frequency (for a dipole interaction $%
V\sim \vec{d}\vec{E}_0$ and $V/\hbar \sim \Omega $ is called the Rabi
frequency). Further study depends on the relationship between frequencies $%
\omega _0$ and $\Omega _0$. Let $\omega _{<}$ and $\omega _{>}$ denote
respectively the smaller and the larger of frequencies $\omega _0\;$and $%
\Omega _0$. Entering a non-dimensional time $t_n=\omega _{>}t$ into
dimensional systems (\ref{eq9}) and (\ref{eq14}) we obtain the following
value $\varepsilon =\omega _{<}/\omega _{>}$.

Three cases are possible:

a) the case of a weak field $\omega _{0}>>\Omega _{0},\,t_{n}=\omega
_{0}t,\,\varepsilon =\Omega _{0}/\omega _{0}$,

b) the case of a strong field $\omega _{0}<<\Omega _{0},\,t_{n}=\Omega
_{0}t,\,\varepsilon =\omega _{0}/\Omega _{0}$,

c) the case in which the frequencies are of the same order, that is, $\omega
_0\sim \Omega _0$. In this case an additional resonance occurs in the system
and the small parameter is absent. This situation requires special
consideration.

Clearly, both $\left[ {E}\right] $ and $\left[ {V}\right] $ are, in general,
functions of $n$ and $m$ which should be taken into account while carrying
out estimates.

11. In problems with initial conditions, the values of the coefficients $%
\left. {c_n}\right| _{t=0}$ in the expansions of the initial functions have
order of unity, whilst those which do not appear $\left( {{\left. {c_n}%
\right| }_{t=0}=0}\right) $ are of order of $\varepsilon $. As follows from
Parseval's equality $\left( \sum\limits_n\left| {c_n}\right| ^2=1\right) $
coefficient $c_n$ must rapidly decreases with the growth of $n$, thus, the
first order approximation in (\ref{eq8}) contains a finite sum with terms
having non-zero coefficients $c_n$ of the expansion of the initial function.

It becomes clear from the above that an accurate account of all possible
internal and external resonances in the system, i.e. the analysis of the
phase relationships, plays a crucial part for obtaining a correct result.
For this reason, it is natural to refer to this perturbation theory as the
phase perturbation theory.

\section{Stationary perturbation of a non-degenerate level of the discrete
spectrum}

   Let us consider a perturbation of a non-degenerate level of the
discrete spectrum, i.e. the case $\hat{V}=\hat{V}(\vec{r})$. The
canonical procedure of averaging is carried out by using
Hamilton's function (\ref
{eq12}) under the condition $\omega _{mn}^{0}\neq O\left( {\varepsilon }%
\right) $ implying no degeneracy and no close energy levels. Simple
calculation by means of eq. (\ref{eq16}) yields
\begin{equation}
\begin{gathered}
\left\{ {H_1 } \right\} = \sum\limits_{n,m} {\frac{{v_{nm} }}
{{\omega _{mn}^0 }}} \bar c_m \bar c_n^ *  \exp ( - i\omega _{mn}^0 t) \hfill \\
\,\,\,\,\,\bar H_1  =  - i\sum\limits_k {v_{kk} \bar c_k \bar c_k^ *  }  \hfill \\
\,\,\,\,\bar H_2  =  - i\sum\limits_{l \ne k} {\sum\limits_k {\frac{{\left| {v_{kl} } \right|^2 }}
{{\omega _{kl}^0 }}\bar c_k \bar c_k^ *  } }  \hfill \\
\end{gathered} \label{eq21}
\end{equation}
  The second approximation to the averaged Hamilton's function (\ref{eq12}) is
given by
\begin{eqnarray}
\bar{H}^{(2)} &=&\varepsilon \bar{H}_{1}+\varepsilon ^{2}\bar{H}%
_{2}=-i\sum\limits_{k}\Delta \omega _{k}\bar{c}_{k}\bar{c}_{k}^{\ast }\,,
\nonumber \\
\Delta \omega _{k} &=&\varepsilon v_{kk}+\varepsilon ^{2}{{\sum\limits_{l}{%
^{\prime }}}}\frac{{{\left| {v_{kl}}\right| }^{2}}}{{\omega _{kl}^{0}}}
\label{eq22}
\end{eqnarray}

Both first and second terms on the averaged Hamilton function can be
renormalised by the phase (frequency) renormalisation in the original
expansion (\ref{eq8}), i.e. by replacing $\omega _k^0$ by $\Omega
_k^0=\omega _k^0+\Delta \omega _k$. In addition to this, $\bar{H}%
^{(2)}\equiv 0$. This procedure can be performed in any order of
calculations. This means in turn that, instead of a standard time interval $%
\Delta t\sim 1/\varepsilon $, this approximation is valid for exponentially
large time intervals which is in full agreement with the general theorems of
mechanics on the behaviour of Hamiltonian systems close to integrable
systems \cite{620}.

The Hamilton function has a diagonal form and the coefficients of the
quadratic form are corrections to the phase (energy) of the unperturbed wave
function.

Equation (\ref{eq20}) of the second approximation for the evolutionary
component $\bar{c}_{k}^{(2)}$ has the form
\begin{equation}
\stackrel{\bullet }{\bar{c}_{k}}=\frac{{\partial \bar{H}^{(2)}}}{{\partial
\bar{c}_{k}^{\ast }}}=-i\Delta \omega _{k}\bar{c}_{k}\;.  \label{eq23}
\end{equation}
Then we easily obtain
\begin{equation}
\bar{c}_{k}^{(2)}=A_{k}\exp (-i\Delta \omega _{k}t)\,.  \label{eq24}
\end{equation}
Coefficients $A_{k}$ are determined from the initial conditions.

The second approximation to coefficients $c_k$ in expansion (\ref{eq8}),
obtained by means of formula (\ref{eq19}), is as follows
\begin{multline}
c_k^{(2)}=A_k^{(2)}\exp (-i\Delta \omega _kt)  \\
+\varepsilon \sum\limits_m\frac{%
{v_{km}}}{{\omega _{mk}^0}}A_m^{(1)}\exp (-i\Omega _{mk}t),  \label{eq25}
\end{multline}
where $\Omega _{mk}=\Omega _m-\Omega _k,\Omega _k=\omega _k^0+\Delta \omega
_k$, and $A_k^{(1)},A_k^{(2)}$ denote the first and the second
approximations to coefficients $A_k$ (it is sufficient to substitute only
the first approximation to $A_k$ into the second term in eq. (\ref{eq25})).

The second approximation to wave function $\Psi ^{(2)}$ is constructed with
the help of coefficients $c_{k}^{(2)}$%

\begin{eqnarray}
\Psi ^{(2)}={{\sum\limits_{k}{^{\prime }}}}\left[ {A_{k}^{(2)}-\varepsilon {%
\sum\limits_{m}{{\frac{{v_{km}}}{{\omega
_{km}^{0}}}}}}A_{m}^{(1)}\exp (-i\Omega _{mk}t)}\right]
\nonumber
\\
\times \Psi _{k}^{0}\exp (-i\Omega _{k}t),  \label{eq26}
\end{eqnarray}

\noindent where $\Omega _{k}=\omega _{k}^{0}+\Delta \omega _{k}$,
i.e. it is sufficient to restrict the consideration by the first
correction to the eigenfrequency.

Provided that the system is in the $n-th$ stationary state of the discrete
spectrum, then $A_{k}=\delta _{kn}$ and we obtain from eq. (\ref{eq26}) that
\begin{equation}
\Psi _{n}^{(2)}=\left[ {\Psi _{n}^{0}+\varepsilon {\sum\limits_{k}{^{\prime }%
{\frac{{v_{kn}}}{{\omega _{nk}^{0}}}}}}\Psi _{k}^{0}}\right] \exp (-i\Omega
_{n}t)\,.  \label{eq27}
\end{equation}

In the case of Cauchy's problem, the system at the initial time instant is
in a certain stationary state of the discrete spectrum, in the $s-th$ state
say, that is $\left. {\Psi (q,t)}\right| _{t=0}=\Psi _{s}^{0}$ and $\left. {%
c{}_{n}}\right| _{t=0}=\delta _{ns}$. Coefficients $A_{n}^{(1)},A_{n}^{(2)}$
are obtained from the relationship
\begin{equation}
\delta _{ns}=A_{n}^{(2)}+\varepsilon \sum\limits_{m}\frac{{v_{nm}}}{{\omega
_{mn}^{0}}}A_{m}^{(1)}\,.  \label{eq28}
\end{equation}
From this relationship we obtain $A_{n}^{(1)}=\delta _{ns}$ and $%
A_{n}^{(0)}=\delta _{ns}-\varepsilon \dfrac{{v_{ns}}}{{\omega _{sn}^{0}}}$.
The second approximation to the wave function has the form
\begin{eqnarray}
\Psi ^{(2)}=\left\{ {\Psi _{s}^{(0)}+\varepsilon {\sum\limits_{k}{^{\prime }{%
\dfrac{{v_{ks}}}{{\omega _{sk}^{0}}}}}}\Psi _{k}^{0}}\right\} \exp
(-i\Omega _{s}t) \nonumber
\\
+\varepsilon
{{\sum\limits_{k}{^{\prime }}}}\dfrac{{v_{ks}}}{{\omega
_{sk}^{0}}}\Psi _{k}^{0}\exp (-i\Omega _{k}t)\,.  \label{eq29}
\end{eqnarray}

This formula is absent in the standard textbooks on quantum mechanics. It is
important to mention that it is adopted in courses on quantum mechanics that
the probability of transition to this problem is determined by the square of
the absolute value of the first correction to the expansion coefficients $%
c_n $, that is, by the second term in eq. (\ref{eq25}), see for example \cite
{613}. The relationships in eq. (\ref{eq26}) show that it is not correct.
This term determines the correction to the unperturbed wave function of the
initial state. The transition probability is determined by the second terms
in $A_n^{(2)}$ which is completely absent in the standard perturbation
theory. This is due to the fact that initial condition $\left. {c_n}\right|
_{t=0}=\delta _{ns}$ is not satisfied. The whole coefficient $c_n^{(2)}$
rather than a part of it, as in the standard theory, must satisfy this
initial condition. In the case under consideration, coincidence is
occasional because the coefficients are independent of time. But these
coefficients are different in the non-stationary theory.

As an example of applying formula (\ref{eq28}), we consider the problem of
the excitation of a charged oscillator by an abruptly applied homogeneous
electric field $\vec{\varepsilon}$, directed along the oscillation axis,
\cite{613}, \cite{626}.

In this case it is necessary to solve the problem
\begin{eqnarray}
i\hbar \frac{\partial \Psi (x,t)}{\partial t} &=&(\hat{H}_{0}+\hat{V})\Psi
(q,t)=(\frac{\hat{p}^{2}}{2m}+\frac{kx^{2}}{2}-e\varepsilon x)\Psi ,
\nonumber \\
\left. {\Psi }\right| _{t=0} &=&\Psi _{0}^{0},\quad \left. {\Psi }\right|
_{x\rightarrow \pm \infty }-\text{bounded},  \label{eq30}
\end{eqnarray}
where $\hat{p}^{2}=-\hbar ^{2}\Delta $, $m$ denotes the oscillator mass, $k$
is the rigidity coefficient, $e$ is the electron charge and $\hat{V}%
=-e\varepsilon x$.

Let us introduce into eq. (\ref{eq30}) a non-dimensional variable $\xi =x/a,$
$(a=\sqrt{{\hbar /m\omega }_0})$ and the eigenfrequency $\omega =\sqrt{k/m}$%
. The eigenfrequency $\omega $ $(\omega =\omega _0)$ is taken as a
characteristic frequency $\omega _0$ and the generalised Rabi frequency is $%
\Omega _0=e\varepsilon a/{\hbar }$. Assuming the external field to be weak,
we enter a small parameter $\varepsilon $ by the relationship $\varepsilon
=\Omega _0/\omega =e\varepsilon a/{\hbar \omega =}e\varepsilon /ka\ll 1$.

An exact solution of the problem of eigenfunctions and eigenvalues in terms
of the non-dimensional units is given by
\begin{eqnarray}
&\Psi _{n}(x,t) =(2^{n}\sqrt{\pi }an!)^{-{{1}/2}}\exp \left[ {-{{\left( {%
\xi -\varepsilon }\right) ^{2}/2}}}\right]\nonumber
\\
&\times H_{n}\left( {\xi -\varepsilon }%
\right) \exp \left[ {-i\left( {n+{{1}/2}-{{\varepsilon
^{2}}}/2}\right) }\right] ,
\nonumber \\
&\omega _{n}=\omega _{0}-\varepsilon ^{2}/2=n+1/2-\varepsilon
^{2}/2\,, \label{eq31}
\end{eqnarray}

\noindent where $H_{n}\left( z\right) $ denotes Hermite
polynomials, \cite{613}, \cite {628}.

The general solution of problem (\ref{eq30}), constructed by means of
eigenfunctions (\ref{eq31}), has the form
\begin{equation}
\Psi (x,t)=\sum\limits_{n=0}^\infty c_n\Psi _n(x,t)\;.  \label{eq32}
\end{equation}
Using the initial condition we obtain the following expansion coefficients $%
c_n$ in eq. (\ref{eq32})

\begin{align*}
c_n={\varepsilon ^n}\left( {2{}^nn!}\right) ^{-1/2}%
{\mathrm{\exp }(-\varepsilon ^2/4)},
\end{align*}

\noindent so that the final result is as follows

\begin{equation}
\begin{gathered}
\Psi (x,t) = \sum\limits_{n = 0}^\infty  {\frac{{\varepsilon ^n {\text{exp}}( - \frac{{\varepsilon ^2 }}
{4})}}
{{\sqrt {2{}^nn!} }}\left( {2^n \sqrt \pi  an!} \right)^{ - {1 \mathord{\left/
{\vphantom {1 2}} \right.
\kern-\nulldelimiterspace} 2}} }  \hfill \\
\;\;\;\;\;\,\,\,\,\,\,\,\,\,\,\,\,\,\,\,\,\,\,\, \times exp\left[ { - {{\left( {\xi  - \varepsilon } \right)^2 } \mathord{\left/
{\vphantom {{\left( {\xi  - \varepsilon } \right)^2 } 2}} \right.
\kern-\nulldelimiterspace} 2}} \right]H_n \left( {\xi  - \varepsilon } \right) \hfill \\
\,\,\,\,\,\,\,\,\,\,\,\,\,\,\,\,\,\,\,\,\,\,\,\,\,\,\,\,\,\,\,\,\,\,\,\,\,\,\,\,\,\,\,\,\,\,\,\,\,\,\,\,\, \times exp\left[ { - i\left( {n + {1 \mathord{\left/
{\vphantom {1 2}} \right.
\kern-\nulldelimiterspace} 2} - {{\varepsilon ^2 } \mathord{\left/
{\vphantom {{\varepsilon ^2 } 4}} \right.
\kern-\nulldelimiterspace} 4}} \right)} \right] \hfill \\
\end{gathered} \label{eq33}
\end{equation}

Performing expansion with accuracy up to terms of order of
$\varepsilon ^2$ yields

\begin{equation}
\begin{gathered}
\Psi ^{(2)} (x,t) = \left\{ {\Psi _0^0  + \varepsilon \frac{1}
{{\sqrt 2 }}\Psi _1^0 } \right. \hfill \\
\left. {\,\,\,\,\,\,\,\,\,\,\,\,\,\,\,\,\,\,\,\,\,\,\,\,\,\,\,\,\,\,\,\,\,\,\,\,\,\,\,\,\,\,\,\,\,\,\,\,\,\,\,\,\,\,\,\,\,\, + \varepsilon \frac{1}
{{\sqrt 2 }}\Psi _1^0 exp( - it)} \right\}exp\left( { - {{it} \mathord{\left/
{\vphantom {{it} 2}} \right.
\kern-\nulldelimiterspace} 2}} \right) \hfill \\
\end{gathered} \label{eq34}
\end{equation}

\noindent where ${\Psi _n^0}$ denotes the eigenfunctions of the
unperturbed problem.

Let us now construct the solution of problem (\ref{eq30}) by means of the
canonical theory of perturbation. The matrix elements of the perturbation
operator are

\[
\begin{gathered}
  \varepsilon v_{mn}  =  - \varepsilon \left[ {\left( {{n \mathord{\left/
 {\vphantom {n 2}} \right.
 \kern-\nulldelimiterspace} 2}} \right)^{ - {1 \mathord{\left/
 {\vphantom {1 2}} \right.
 \kern-\nulldelimiterspace} 2}} \delta _{m,n - 1} } \right. \hfill \\
  \,\,\,\,\,\,\,\,\,\,\,\,\,\,\,\,\,\,\,\,\,\,\,\,\,\,\,\,\,\,\,\,\,\,\,\,\,\,\,\,\,\,\,\,\,\,\,\,\,\,\,\,\,\,\left. { + \left( {{{\left( {n + 1} \right)} \mathord{\left/
 {\vphantom {{\left( {n + 1} \right)} 2}} \right.
 \kern-\nulldelimiterspace} 2}} \right)^{ - {1 \mathord{\left/
 {\vphantom {1 2}} \right.
 \kern-\nulldelimiterspace} 2}} \delta _{m,n + 1} } \right] \hfill \\
\end{gathered}
\]

\noindent the corrections to the eigenfrequencies are $\Delta
\omega _k=-\varepsilon ^2/2$, the eigenfrequencies of the
unperturbed system are $\omega _0^0=1/2,$ $\omega _1^0=3/2,$ and
eq. (\ref{eq29})\ takes the form

\[
\begin{gathered}
   \hfill \\
  \Psi ^{(2)} (x,t) = \left\{ {\Psi _0^0  + \varepsilon \frac{1}
{{\sqrt 2 }}\Psi _1^0 } \right. \hfill \\
  \left. {\,\,\,\,\,\,\,\,\,\,\,\,\,\,\,\,\,\,\,\,\,\,\,\,\,\,\,\,\,\,\,\,\,\,\,\,\,\,\,\,\,\,\,\,\,\,\,\, + \varepsilon \frac{1}
{{\sqrt 2 }}\Psi _1^0 exp( - it)} \right\}exp\left( { - {{it} \mathord{\left/
 {\vphantom {{it} 2}} \right.
 \kern-\nulldelimiterspace} 2}} \right) \hfill \\
\end{gathered}
\]

As expected, the solution constructed by the perturbation theory coincides
with the series expansion of the exact solution. Let us notice that
obtaining solution (\ref{eq34}) is simpler than expanding the exact solution
in the series.

The probability of transition to the first excitation state is the square of
the absolute value of the coefficient of ${\Psi _1^0\exp (-i3t/2)}$ and is
equal to $\varepsilon ^n/2$.

\section{Stationary excitation of two close levels}

In the case of stationary perturbation, it is expedient to take into account
the principal resonance by renormalisation of the frequency in the original
expansion, i.e. to present expansion (\ref{eq8}) as follows
\begin{equation}
\Psi (q,t)=\sum\limits_{n}c_{n}(t)\psi _{n}^{0}(q)\exp (-i\Omega _{n}^{0}t),
\label{eq35}
\end{equation}
where $\Omega _{n}^{0}=\omega _{n}^{0}+\varepsilon v_{nm}$. This
transformation removes the term with $\varepsilon v_{nm}$ from the equations
and the effective Hamilton function, the latter taking the form
\begin{equation}
\varepsilon H_{1}(c,c^{\ast },t)=-i\varepsilon {\sum\limits_{n,m}{^{\prime }}%
}v_{nm}c_{m}c_{n}^{\ast }\exp (i\Omega _{nm}^{0}t)\,.  \label{eq36}
\end{equation}
The presence of two close levels in the unperturbed system, for example with
indices $\alpha $ and $\beta $ ($\omega _{\alpha }^{0}-\omega _{\beta
}^{0}=\varepsilon \delta _{0})$ means that $\Omega _{\alpha \beta
}^{0}=\Omega _{\alpha }^{0}-\Omega _{\beta }^{0}=\varepsilon \delta $ $%
(\delta =\delta _{0}+v_{\alpha \alpha }-v_{\beta \beta })$ and leads to the
necessity to take into account the dependence of sum (\ref{eq36}) on slow
time $\tau =\varepsilon t$ in the process of averaging
\begin{eqnarray}
\varepsilon H_{1} =&-i\varepsilon v_{\alpha \beta }c_{\alpha
}^{\ast }c{}_{\beta }\exp (i\delta \tau )-i\varepsilon v_{\beta
\alpha }c_{\alpha }c_{\beta }^{\ast }\exp (-i\delta \tau
)\nonumber \\&-i\varepsilon {\sum\limits_{n,m}{^{\prime \prime
}}}v_{nm}c_{m}c_{n}^{\ast }\exp (-i\Omega _{nm}^{0}t),
\label{eq37}
\end{eqnarray}
where notation $\sum^{\prime \prime }$ implies that this sum does not
contain the diagonal elements ($m=n)$ and the elements with $m=\alpha
,n=\beta \;$and $m=\beta ,n=\alpha $.

The calculation of the first-order approximation yields the following
averaged Hamilton function
\begin{multline}
\varepsilon \bar{H}_{1}=-i\varepsilon v_{\alpha \beta }\bar{c}_{\alpha
}^{\ast }\bar{c}{}_{\beta }\exp (i\delta \tau ) \\ -i\varepsilon v_{\beta \alpha
}\bar{c}_{\beta }^{\ast }\bar{c}_{\alpha }\exp (-i\delta \tau )\,.
\label{eq38}
\end{multline}

The Hamilton equations for the evolutionary components $\bar{c}_{\alpha },%
\bar{c}_{\beta }$ have the form
\begin{eqnarray}
\dot{\bar{c}}_{\alpha } &=&\varepsilon \frac{{\partial \bar{H}_{1}}}{{%
\partial \bar{c}_{\alpha }^{\ast }}}=-i\varepsilon v_{\alpha \beta }\bar{c}%
_{\beta }\exp (i\delta \tau )\,,  \nonumber \\
\dot{\bar{c}}_{\beta } &=&\varepsilon \frac{{\partial \bar{H}_{1}}}{{%
\partial \bar{c}_{\beta }^{\ast }}}=-i\varepsilon v_{\beta \alpha }\bar{c}%
_{\alpha }\exp (-i\delta \tau )\,.  \label{eq39}
\end{eqnarray}

Thus, the averaging procedure ''cuts'' a two-level system with close levels $%
\alpha $ and $\beta $ from the whole spectrum. It is evident that the case
of two-fold degeneracy is a particular case of this problem for $\delta
_{0}=0$. The same solution has the problem of resonant interaction with an
external field with frequency $\omega \;(\omega -\omega _{\alpha \beta
}^{0}=\varepsilon \delta _{0})$, the difference being only in the resonance
type. In the problem of the close levels there is an internal resonance.

The solution of problem (\ref{eq39}) subject to the initial condition $%
\left. {\bar{c}_{\alpha }}\right| _{t=0}=\delta _{n\alpha }$ is as follows
\begin{eqnarray}
\bar{c}_{\alpha } &=&\frac{{1}}{{\Delta }}\left[ {\Omega _{1}\exp (i\Omega
_{2}\tau )-\Omega _{2}\exp (i\Omega _{1}\tau )}\right] ,  \nonumber \\
\bar{c}_{\beta } &=&\frac{{g^{\ast }}}{{\Delta }}\left[ {\exp (-i\Omega
_{1}\tau )-\exp (-i\Omega _{2}\tau )}\right] ,  \nonumber \\
\bar{c}_{n} &=&0,\;\;n\neq \alpha ,\beta \;,  \label{eq40}
\end{eqnarray}
where $\Omega _{1,2}=\dfrac{{\delta }}{{2}}\pm \dfrac{{\Delta }}{{2}},$ $%
\Delta =\sqrt{\delta ^{2}+4{\left| {g}\right| }^{2}},$ $g=v_{\alpha \beta }$.

Representing the first equation in eq. (\ref{eq40}) in the form
\begin{equation}
\bar{c}_{\alpha }=\frac{{\exp (i\Omega _{1}\tau )}}{{\Delta }}\left\{ {%
\Delta +\Omega _{1}[\exp (-i\Delta \tau )-1]}\right\} ,  \label{eq41}
\end{equation}
it is easy to find that the probability of being in the state become unity $%
\left( w_{\alpha }=1\right) $ after the time interval $\tau ^{\ast }=2\pi
/\Delta $. Thus, the oscillations between the levels $\alpha $ and $\beta $
have period $T=2\pi /\Delta $ or frequency $\omega =\Delta $.

Substituting coefficients (\ref{eq40}) into expansion (\ref{eq35}) we obtain
that the wave function is a superposition of two stationary states with
\begin{equation}
\omega _{\alpha ,\beta }=\frac{{\Omega _{\alpha }^{0}+\Omega _{\beta }^{0}}}{%
{2}}\pm \frac{{1}}{{2}}\sqrt{\left( {\Omega _{\alpha }^{0}-\Omega _{\beta
}^{0}}\right) ^{2}+4{\left| {g}\right| }^{2}}\,.  \label{eq42}
\end{equation}

In the case of a degenerate level $\delta _{0}=0$ and from eq. (\ref{eq42})
we obtain the correction $\Delta \omega $ to the frequency of the stationary
state $\omega _{\alpha }^{0}=\omega _{\beta }^{0}=\omega ^{0}$%
\begin{equation}
\Delta \omega =\varepsilon \frac{{v_{\alpha \alpha }+v_{\beta \beta }}}{{2}}%
\pm \frac{{1}}{{2}}\sqrt{\left( {v_{\alpha \alpha }-v_{\beta \beta }}\right)
^{2}+4{\left| {v_{\alpha \beta }}\right| }^{2}}\,.  \label{eq43}
\end{equation}

Therefore, all three problems, namely the problems on close levels, two-fold
degenerate level and resonant interaction with an external single-frequency
field, are all solved in the framework of the same approach and yield the
results coinciding with the traditional one with first order accuracy.

While solving the problem, we determine the conditions under which the
quantum system with a discrete spectrum can be modelled, in the first
approximation, by a two-level system. The main point of this procedure is
the possibility of averaging Hamilton's function (\ref{eq38}). The condition
of weakness of the external field is needed for this. Then, the non-trivial
initial conditions are required, at least for one of the coefficients $%
c_\alpha ,c_\beta $, otherwise the solutions of the homogeneous equations in
(\ref{eq39}) are trivial.

Nowadays, the procedure of solving these problems is performed backwards. A
two-level systems is first taken, then a so-called resonant approximation
(rotating wave approximation), \cite{613}, \cite{616}, is applied to it. At
this stage it is incorrectly assumed that the resonant approximation is also
applicable in the cases where the perturbation theory is invalid, see for
example \cite{616}.

\section{Non-stationary Schr\"{o}dinger's equation as a Hamiltonian system}

The situation
\index{Schr\"{o}dinger's equation!non-stationary}studied in the
non-stationary perturbation theory%
\index{non-stationary perturbation theory} occurs when operator $%
\hat{H}$ can be represented as sum $\hat{H}=\hat{H}_{0}+\varepsilon \hat{V}%
(q,t)$ $(0<\varepsilon <<1)$ of two self-adjoint operators. In the
\index{adiabatic approximation}adiabatic approximation the perturbation
operator $%
\hat{V}(q,t)$ is not small and depends on slow time $\tau =\varepsilon t$
such that $\hat{V}(q,t)=\hat{V}(q,\tau )$. The solution should be
constructed within the asymptotically large time interval $\tau \sim
1/\varepsilon $ when change in the perturbation operator is large. In this
case splitting the total Schr\"{o}dinger's operator $\hat{H}$ into two
operators, namely the generating (unperturbed) operator and a perturbation
operator makes no sense. In order to embrace both possibilities we consider
problem (\ref{eq4}) with the time-dependent Schr\"{o}dinger operator $\hat{H}%
=\hat{H}(q,t)$.

Let us assume that the stationary problem corresponding to (\ref{eq4}) is
solvable for a parametric dependence of Schr\"{o}dinger's operator on time
and has a discrete spectrum. This means that the eigenfunctions and the
eigenvalues of the problem are given by
\begin{equation}
\hat{H}\left( t\right) \psi _n(q,t)=E_n(t)\psi _n(q,t),  \label{eq44}
\end{equation}
with time $t$ being fixed. The eigenfunctions are assumed to be
orthonormalised as follows
\begin{equation}
\int\limits_{-\infty }^\infty \bar{\psi}_m(q,t)\psi _n(q,t)dq=\delta _{mn}
\label{eq45}
\end{equation}
where a bar denotes the complex conjugate.

The existing\ approximations of Born-Fock \cite{628} and Landau-Dykhne \cite
{615}, \cite{616} suggest that the eigenfunctions can be chosen as being
real-valued (i.e. no magnetic field is assumed) which essentially reduces
the applicability of the method. In the present study this assumption is not
needed.

In the case of weak fields the results of the adiabatic approximation of
Landau-Dykhne do not coincide with the results of perturbation theory \cite
{616}. The approximation of Born-Fock is actually not an approximation at
all since all higher approximations turn out to be of the order of the first
approximation \cite{616}. In addition to this, both approximations yield an
incorrect factor in front of the exponential function \cite{616}.

Let us look for the solution of the exact problem in the form
\begin{equation}
\Psi (q,t)=\sum\limits_{n}c_{n}(t)\psi _{n}(q,t)\exp \left\{ {-i{%
\int\limits_{0}^{t}{\Omega _{n}(z)dz}}}\right\} ,  \label{eq46}
\end{equation}
where

\begin{eqnarray*}
\Omega _{n}(t)&=&\omega _{n}(t)+v_{nn}(t),\;  \\ \omega
_{n}(t)&=&E_{n}(t),\;  \qquad  v_{nn}=-i\int\limits_{-\infty }^{\infty }\bar{\psi}_{n}%
\frac{{\partial \psi _{n}}}{{\partial t}}dq.
\end{eqnarray*}

The meaning of this choice of the phase becomes clear in what
follows.

Inserting eq. (\ref{eq46}) into eq. (\ref{eq4}) yields the following
equation for the expansion coefficients $c_{m}(t)$%
\begin{equation}
\dot{c}_{m}(t)=-i{\sum\limits_{n,m}{^{\prime }}}v_{mn}c_{n}\exp \left\{ {-i{%
\int\limits_{0}^{t}{\Omega _{mn}(z)dz}}}\right\} ,  \label{eq47}
\end{equation}
where a dot denotes a total time derivative and a prime at the summation
sign denotes the absence of a diagonal components with $m=n$. The matrix of
coefficients $v_{mn}$ has the form
\begin{equation}
v_{mn}=-i\int\limits_{-\infty }^{\infty }\bar{\psi}_{m}(q,t)\frac{{\partial
\psi _{n}(q,t)}}{{\partial t}}dq  \label{eq48}
\end{equation}
and is Hermitian, i.e. $v_{mn}=\bar{v}_{nm}$.

The choice of phase indicated in eq. (\ref{eq46}) ensures that the sum has
no diagonal component which is responsible for the principal resonance. If
eq. (\ref{eq46}) had this diagonal component, the sum (\ref{eq47}) would
have a small resonant denominator.

Indeed, differentiating eq. (\ref{eq44}) with respect to time and taking
into account that Schr\"{o}dinger's operator is self-adjoint, we obtain
\begin{equation}
v_{mn}=\left( \frac i{\hbar \omega _{mn}}\right) \left( \frac{\partial \hat{H%
}}{\partial t}\right) _{mn},\quad m\neq n\;.  \label{eq49}
\end{equation}
It is clear that in the case of the real-valued eigenfunctions $\bar{\psi}%
_n=\psi _n$, that is, the diagonal elements $v_{nn}=0$. It is this fact that
is the reason for the real-valued normalisation in the Born-Fock
approximation.

Similar actions in the case when $m=n$ leads to the relationship $\left(
\dfrac{\partial \hat{H}}{\partial t}\right) _{nn}$ $=\dfrac{\partial E_{n}}{%
\partial t}$ and do not determine the diagonal matrix elements. In the case
in which Schr\"{o}dinger's operator depends on time $\tau $ in terms of the
set of functions $\xi _{i}(\tau )$ $\left( i=1,2..N\right) ,$ elements $%
v_{nn}$ determine the topological adiabatic Berry phase \cite{629} whose
value does not depend on the evolution time and is determined only by a
closed contour in the parameter space.

Equation (\ref{eq49}) indicates three cases allowing the development of the
perturbation theory. In the adiabatic case $\hat{H}=\hat{H}(\xi (\tau )),$
so that $\dfrac{\partial \hat{H}}{\partial t}=\varepsilon \left( \dfrac{{%
\partial }\hat{H}}{\partial \xi }\right) \left( \dfrac{{\partial \xi }}{{%
\partial \tau }}\right) $. In the case of the non-stationary perturbation
theory, Schr\"{o}dinger's operator has the form $\hat{H}=\hat{H}%
_0+\varepsilon \hat{V}(q,t)$. Finally, in the case of the adiabatic
perturbation theory $\hat{H}=\hat{H}_0+\varepsilon \hat{V}(q,\tau )$.

The system of equations (\ref{eq47}), along with the complex conjugate one,
is Hamiltonian (in the classical sense) having the following Hamilton
function
\begin{multline}
H(c,c^{\ast },t)=-i{\sum\limits_{n,m}{^{\prime }}}v_{mn}(t)c_{n}  \\
\times c_{m}^{\ast}\exp \left\{ {-i{\int\limits_{0}^{t}
{\Omega _{mn}(z)dz}}}\right\}\label{eq50}
\end{multline}
which describes the classical distributed system. The matrix of coefficients
$v_{mn}$ is Hermitian which ensures that this system is Hamiltonian and in
turn enables one to apply the phase perturbation theory.

\section{Adiabatic approximation}

We assume that Schr\"{o}dinger's operator has the form $\hat{H}=\hat{H}_{0}+%
\hat{V}(q,\xi (\tau )),$ where $\tau =\varepsilon t$ denotes slow time. Then
the matrix elements $v_{mn}\sim \varepsilon \xi ^{\prime }$ and Hamilton's
function (\ref{eq50}) can be cast in the form
\begin{multline}
\varepsilon H_{1}(c,c^{\ast },t,\tau )=-i\varepsilon {\sum\limits_{n,m}{%
^{\prime }}}v_{mn}(\tau )c_{n}  \\  \times c_{m}^{\ast }\exp \left\{ {i{%
\int\limits_{0}^{t}{\Omega _{mn}(\tau )dt}}}\right\} \,,  \label{eq51}
\end{multline}
where $\dot{\tau}=\varepsilon $ and $\Omega _{mn}=\Omega _{m}-\Omega _{n}$.

The canonical form allows us to convert the evolutionary equations by means
of the formulae
\begin{gather}
\bar{H}^{(2)}(\bar{c},\bar{c}^{\ast },\tau )=\varepsilon \bar{H}_{1}(\bar{c},%
\bar{c}^{\ast },\tau )+\varepsilon ^{2}\bar{H}_{2}(\bar{c},\bar{c}^{\ast
},\tau )\,,  \nonumber \\
\bar{H}_{1}=\langle H_{1}\rangle ,\;\bar{H}_{2}=-\left\langle {\left( \frac{{%
{{\partial \tilde{H}_{1}}}}}{{\partial }\bar{c}^{\ast }}\right) \left( \frac{%
{{{\partial {\left\{ {H_{1}}\right\} }}}}}{{\partial }\bar{c}}\right) }%
\right\rangle \,,  \label{eq52}
\end{gather}
where $\bar{H}^{(2)}$ denotes the second approximation to the averaged
Hamilton's function, whilst $\bar{c}=\left( {\bar{c}_{1},\bar{c}_{2},..}%
\right) $ and $\bar{c}^{\ast }=\left( {\bar{c}_{1}^{\ast },\bar{c}_{2}^{\ast
},..}\right) $ are the evolutionary components of variables $c$ and $c^{\ast
}$.

Averaging expression (\ref{eq51}) along the generating solution $(c_k  = const,\tau  = const)$
, we obtain $\bar H_1  = \langle H_1 \rangle  = 0$
 which in turn implies that $\mathop {\bar c_k }\limits^ \bullet   = 0$.
  The latter result is the adiabatic theorem of Kato
\cite{612} which is obtained in fact without calculations, cf.
(\cite{630}) for the proof. In the classical sense, the
evolutionary components $\bar{c}_{k}$ of the original variables
$c_{k}$ are the adiabatic invariants, see \cite{608}, \cite{620},
i.e. they retain the initial values for the asymptotic time
interval $t\sim 1/\varepsilon $. For deriving this result it is
necessary to assume that $\Omega _{mn}(\tau )\neq O(\varepsilon
)$, i.e. the system has no degeneracy, there are no close levels
and the levels do not intersect during the evolution time.

In the
\index{adiabatic approximation}adiabatic (first) approximation, the solution
of Schr\"{o}dinger's equation has the form
\begin{multline}
\Psi ^{(1)}(q,t)=\sum\limits_{n=0}^{\infty }c_{n}^{(1)}\psi _{n}(q,t)  \\
\times\exp\left\{ {-i{\int\limits_{0}^{t}{\Omega _{mn}(z)dz}}}\right\} \;.
\label{eq53}
\end{multline}

Under rather general assumptions, Los's theorem \cite{618} renders estimates
for the difference $\left| {{\left. {\Psi (q,t)-\Psi ^{(1)}(q)}\right| }}%
\right. <C\varepsilon $, where $C$ is a constant independent of $\varepsilon
$ for time interval $t\sim 1/\varepsilon $.

\section{Post-adiabatic approximation}

In order to construct the second (post-adiabatic) approximation we make use
of relationships in eq. (\ref{eq52}). Simple calculation yields
\begin{equation}
\bar{H}_{2}=-i\sum\limits_{k}\Delta \Omega _{k}(\tau )\bar{c}_{k}\bar{c}%
_{k}^{\ast },\;\Delta \Omega _{k}={\sum\limits_{l}{^{\prime }}}\frac{{{%
\left| {v_{kl}}\right| }^{2}}}{{\Omega _{kl}}}\,,  \label{eq54}
\end{equation}
so that the second approximation to the averaged Hamilton's function $\bar{H}%
^{(2)}$ has the form
\begin{equation}
\bar{H}^{(2)}=-i\varepsilon ^{2}\sum\limits_{k}\Delta \Omega _{k}(\tau )\bar{%
c}_{k}\bar{c}_{k}^{\ast }\,.  \label{eq55}
\end{equation}

Hamiltonian equations with Hamilton's function (\ref{eq55}) for the
evolutionary components $\bar{c}_{k}$ are integrated easily, to give
\begin{multline}
\bar{c}_{k}=A_{k}\exp \left\{ -i\varepsilon ^{2}\int\limits_{0}^{\tau
}\Delta \Omega _{k}(z)dz\right\} \\
=A_{k}\exp \left( -i\alpha _{k}\right) \,.\label{eq56}
\end{multline}
Integration constants $A_{k}$ are determined by means of the initial
conditions.

Let us notice that the phase of coefficients $\bar{c}_{k}$ could be included
into the original expansion (\ref{eq46}), then we would obtain $\bar{H}%
_{2}=0,$ $\bar{H}^{(2)}=0$.

The second approximation to the expansion coefficients in eq. (\ref{eq46})
is constructed by means of formulae (\ref{eq16})
\begin{multline}
c_{k}^{(2)}=A_{k}^{(2)}\exp (-i\alpha _{k}) \\
-\varepsilon {\sum\limits_{m}{%
^{\prime }}}\frac{v_{km}}
{\Omega _{km}}A_{m}\exp (-i\alpha _{m}+i\Omega
_{km}t),  \label{eq57}
\end{multline}
With the help of coefficients $c_{k}^{(2)}$ we obtain the second
approximation $\Psi ^{(2)}(q,t)$ to the solution of Schr\"{o}dinger's
equation
\begin{widetext}
\begin{equation}
\Psi ^{(2)}(q,t)=\sum\limits_{k}\biggl[ A_{k}{\exp }(-i\alpha
_{k})-
\varepsilon {\sum\limits_{m}{^{\prime }}}\dfrac{v_{km}}{\Omega _{km}}%
A_{m}\exp (-i\alpha _{m}-i\Omega _{km}t)\biggr] \Psi _{k}\exp
\left[ -i\int\limits_{0}^{t}\Omega _{k}(z)dz\right] \,.
\label{eq58}
\end{equation}
\end{widetext}
It is necessary to mention that in eq. (\ref{eq57}) we can limit
our
consideration to the terms in the sum by the first approximation $%
A_{m}^{(1)} $ with respect to $\varepsilon .$

When Cauchy's problem is studied, the system at the initial time instant is
at a certain stationary state, say $s-th$, of the discrete spectrum of the
unperturbed problem with Schr\"{o}dinger's operator $\hat{H}_{0}$, i.e. $%
\left. {\Psi (q,t)}\right| _{t=0}=\Psi _{s}^{0}$. In contrast to the
stationary case we can not take that $\left. {c_{n}}\right| _{t=0}=\delta
_{ns}$, since the expansion is carried out in terms of the eigenfunctions of
the perturbed problem $\psi _{n}(q,t)=\psi _{n}(q,\xi (t))$, where $\xi (t)$
denotes parameters determining the dependence of the perturbation on time.
With this in view, we additionally assume that the switch fulfills the
conditions $\xi (0)=\dot{\xi}(0)=0$. The problem can be solved under other
conditions which implies an instantaneous switching of the perturbation
followed by its adiabatic change.

Therefore, we take that the equation for coefficients $A_{k}$ has the
following form
\begin{equation}
\left. {c{}_{k}}\right| _{t=0}=\delta _{ks}=A_{k}-\varepsilon {%
\sum\limits_{m}{^{\prime }}}\left[ {{\frac{{v_{km}(\tau )}}{{\Omega
_{km}(\tau )}}}}\right] _{t=0}A_{m}\,.  \label{eq59}
\end{equation}
Obviously the matrix elements $v_{km}\sim \dot{\xi}(\tau )$, thus under the
adopted conditions we obtain $A_{ks}=\delta _{ks}$, that is
\begin{equation}
c_{ks}^{(2)}=\delta _{ks}\exp (-i\alpha _{k})-\varepsilon \frac{{v_{ks}}}{{%
\Omega _{ks}}}\exp (-i\alpha _{s}+i\Omega _{ks}t)\,.  \label{eq60}
\end{equation}
Finally, the second (post-adiabatic) approximation $\Psi ^{(2)}$ for the
wave function is as follows
\begin{equation}
\begin{gathered}
\Psi ^{(2)} (q,t) = \sum\limits_k {\left[ {A_k \exp ( - i\alpha _k )} \right.}  \hfill \\
\left. {\,\,\,\,\,\,\,\,\,\,\,\,\,\,\,\,\,\,\,\, - \varepsilon \sum\limits_m {'\frac{{v_{km} }}
{{\Omega _{km} }}} A_m exp( - i\alpha _m  - i\Omega _{km} t)} \right] \hfill \\
\,\,\,\,\,\,\,\,\,\,\,\,\,\,\,\,\,\,\,\,\,\,\,\,\,\,\,\,\,\,\,\,\,\,\,\,\,\,\,\,\,\,\,\,\,\ \times \Psi _k exp\left[ { - i\int\limits_0^t {\Omega _k } (z)dz} \right] \hfill \\
\end{gathered} \label{eq61}
\end{equation}
 It can be proved, see \cite{618}, that the estimate $\left| {{{\Psi
(q,t)-\Psi ^{(2)}(q,t)}}}\right| <B\varepsilon ^{2}$ is valid for the time
interval $t\sim 1/\varepsilon $. Papers by physicists do not take into
account the boundedness of the time interval in which an approximate
solution approximates the exact solution. As one can see from the
forthcoming examples, this interval, in general, can not be enlarged.

The problem of constructing the adiabatic approximation and taking account
of the transition through the virtual levels was posed by Dykhne: ''In order
to obtain the correct factor of the exponential function it would be
necessary to take into account all higher approximations of the perturbation
theory, all yielding results of the same order. In practice, this is, of
course, not feasible. The obtained formulae give answers to the question of
calculating the probability of transition of a quantum system to an
''adjacent'' level. As for transitions to more remote levels, then the
transitions through virtual levels may compete with the considered process
of the ''direct'' transition. However, this question needs an additional
investigation'', \cite{615}. Nevertheless, this problem has not been solved
so far by the existing methods of perturbation theory.

\section{Quantum linear oscillator in a variable homogeneous field}

In order to compare the obtained results with the known ones, let us study
the following problem having an exact solution. The situation considered is
the motion of a particle in the field of a parabolic potential subjected to
a variable external force, i.e.

\begin{equation}
\begin{split}
&i\hbar \frac{{\partial \Psi (q,t)}}{{\partial t}}=\left[ {-{\frac{{\hbar ^2}%
}{{2m}}}{\frac{{\partial ^2}}{{\partial q^2}}}+{\frac{{m\omega ^2q^2}}{{2}}}%
-e\varepsilon _0f(\nu t)q}\right] \Psi (q,t)
\\
\label{eq62}&\left. {\Psi (q,t)}\right| _{t=0}=\pi
^{-{1/4}}a^{{1/2}}\exp \left[ {-\left( {q/a}\sqrt{2}\right)
^2}\right] ,\\
&\left.{\Psi (q,t)}\right| _{q\rightarrow \pm \infty
}-\text{bounded},
\end{split}
\end{equation}

\noindent where $m$ and $e$ denote the mass and charge of the oscillator, ${%
\varepsilon _0}$ is amplitude of the electric field, $\nu ^{-1}$ denotes a
characteristic time constant of the field, $a=\left( {\hbar /m\omega }%
\right) ^{{1/2}}$ denotes a characteristic length scale. Clearly, the
initial state is the main state of the free harmonic oscillator. Let us
introduce the non-dimensional time $t_n=\omega t$, the non-dimensional
coordinate $x=q/a$ and the non-dimensional force amplitude $\varepsilon
_1=e\varepsilon _0a/\hbar \omega =\Omega /\omega $. Problem (\ref{eq62}) in
the terms of non-dimensional variables takes the form

\begin{equation}
\begin{split}
&i\frac{{\partial \Psi (x,t)}}{{\partial t}} =\left[ {-{\frac{{1}}{{2}}}{%
\frac{{\partial ^2}}{{\partial x^2}}}+{\frac{{1}}{{2}}}x^2-\xi
(\varepsilon t)x}\right] \Psi (x,t),
\\
\label{eq63}&\left. {\Psi (x,t)}\right| _{t=0} =\pi
^{-{{1}/4}}\exp \left( {-{x^2}/2}%
\right) , \\
&\left. {\Psi (x,t)}\right| _{x\rightarrow \pm \infty
}-\text{bounded\thinspace ,}
\end{split}
\end{equation}

where $\varepsilon =\nu /\omega $, $\xi (\varepsilon t)=\varepsilon
_1f(\varepsilon t)$ and the non-dimensional time is denoted by $t$.

The exact solution of this problem is given by, cf. \cite{630},

\begin{eqnarray}
\Psi _0(x,t) &=&\pi ^{-{1/4}}\exp \left\{ {-{\dfrac{{it}}{{2}}}+i{%
\int\limits_0^t{dz\delta ^2(z)\exp (-2iz)}}-}\right.
\nonumber \\
&&\left. {{\frac{{x^2}}{{2}}}-\sqrt{2}x\delta (t)\exp
(-it)}\right\},\label{eq64}
\end{eqnarray}

\begin{equation}
\delta (t)=-\frac{{i}}{\sqrt{2}}\int\limits_0^t\xi (\varepsilon z)\exp (iz)dz
\label{eq65}
\end{equation}

for arbitrary values of the parameters $\varepsilon $ and $\varepsilon _1$.

In order to construct expansion (\ref{eq64}) in the adiabatic case $\left( {%
0<\varepsilon <<1}\right) $ we integrate eq. (\ref{eq65}) by parts three
times and take into account that $\xi (0)={{\dot{\xi}}}(0)=\ddot{\xi}(0)=0$.
The result is
\begin{equation}
\delta (t)=-\frac{{1}}{\sqrt{2}}\left[ {\xi +i}\xi {-{\dot{\xi}}}\right]
\exp (it)-\frac{{i}}{{2}}\int\limits_{0}^{t}\exp (iz)\dddot{\xi}(z)dz\,.
\label{eq66}
\end{equation}
Taking $M=\max \left| \ddot{\xi}{(t)}\right| $ in time interval $\left[ {0,T}%
\right] $ we obtain the following estimate
\begin{equation}
\left| {{\int\limits_{0}^{T}{\exp (iz)}}}\dddot{\xi}{(z)dz}\right| \leqslant
\int\limits_{0}^{T}\left| \dddot{\xi}{(z)}\right| dz\leqslant MT\sim
\varepsilon ^{3}T\,.  \label{eq67}
\end{equation}
It is clear from this equation that one can neglect the latter term in eq. (%
\ref{eq66}) within time interval $\Delta t\sim T\sim 1/\varepsilon $. For
this reason, it is necessary to keep the second order term $\ddot{\xi}$ in
eq. (\ref{eq66}) for evaluating integral (\ref{eq64}) and omit it by
substituting without integration.

With the same accuracy we evaluate the following integral
\begin{equation}
i\int\limits_{0}^{t}dz\exp (-2iz)\delta ^{2}(z)=i\int\limits_{0}^{t}\frac{{%
\xi ^{2}}}{{2}}dz-\frac{{\xi ^{2}}}{{2}}-i\xi {{\dot{\xi}}}%
+i\int\limits_{0}^{t}\frac{\dot{\xi}^{2}}{{2}}dz  \label{eq68}
\end{equation}
and the expansion of the exact solution (\ref{eq64}) as a series in time
interval $t\sim 1/\varepsilon $%
\begin{eqnarray}
\Psi _{0}(x,t) &=&\left[ {\Psi _{0}(x-\xi )+i}\dot{\xi}{\Psi _{1}(x-\xi )}%
\right] \exp \left\{ {-{\dfrac{{it}}{{2}}}+}\right.  \nonumber \\
&&\left. {i{\int\limits_{0}^{t}{{\frac{{\xi ^{2}}}{{2}}}dz}}+i{%
\int\limits_{0}^{t}{{\frac{\dot{\xi}^{2}}{{2}}}dz}}}\right\} +O\left( {%
\varepsilon ^{2}}\right) \,.  \label{eq69}
\end{eqnarray}

Next, we construct the expansion of the exact solution (\ref{eq64}) in the
case of the harmonic non-resonant perturbation $(\xi \left( {t}\right)
=\varepsilon _{1}\sin \nu t,0<\varepsilon _{1}<<1,\;\nu \neq O\left( {%
\varepsilon _{1}}\right) ,\;\nu \neq 1).$ In this case
\begin{equation}
\delta (t)=-\varepsilon _{1}\frac{{i}}{{\sqrt{2}(\nu ^{2}-1)}}\left[ {\left(
{i\sin \nu t-\nu \cos \nu t}\right) \exp (it)+\nu }\right]  \label{eq70}
\end{equation}
\begin{equation}
i\int\limits_{0}^{t}dz\delta ^{2}(z)\exp (-2it)=-i\frac{{\varepsilon _{1}^{2}%
}}{{4(\nu ^{2}-1)}}t+O\left( {\varepsilon _{1}^{2}}\right)  \label{eq71}
\end{equation}

The terms omitted in eq. (\ref{eq71}) are uniformly bound such that
approximation (\ref{eq71}) is valid at any time instant. Inserting eqs. (\ref
{eq70}) and eq. (\ref{eq71}) into eq. (\ref{eq64}) yields the following
expansion of the exact solution

\begin{widetext}
\begin{eqnarray}
\Psi _0(x,t)=\pi ^{-{1/4}}\exp \left\{ {-{\frac{{it}}{{2}}}-i{\frac{{%
\varepsilon _1^2}}{{4(\nu ^2-1)}}}t-{\frac{{x^2}}{{2}}}+x{\frac{{%
i\varepsilon _1}}{{\nu ^2-1}}}}\left( {i\sin \nu t-}\right.
\right. \left. \left. {\nu \cos \nu t}\right) {+\exp
(-it){\dfrac{{i\varepsilon
_1\nu }}{{\nu ^2-1}}}x}\right\}= \nonumber
\end{eqnarray}
\[
\begin{gathered}
  \Psi ^{(2)} (x,t) = \left\{ {\left[ {\Psi _0  + \frac{{i\varepsilon _1 }}
{{\sqrt 2 (\nu ^2  - 1)}}\Psi _1 (i\sin \nu t - \nu \cos \nu t)} \right]\exp \left( { - \frac{{it}}
{2}} \right)} \right. \hfill \\
  \;\;\;\;\;\;\;\;\;\;\;\;\;\;\;\;\;\;\;\;\;\;\;\,\,\,\,\,\,\,\,\,\,\,\,\,\,\,\,\,\,\,\,\,\,\,\,\,\,\,\,\,\,\,\,\,\,\,\,\,\,\,\,\,\,\,\,\,\,\,\,\,\,\,\,\,\,\,\,\,\,\,\,\,\,\,\,\,\,\,\,\,\,\,\,\,\,\,\,\,\,\,\,\,\,\,\,\,\,\,\,\,\,\,\,\,\,\,\,\,\,\,\,\,\,\,\,\,\,\,\,\,\,\,\,\,\,\;\left. { + \frac{{i\varepsilon _1 \nu }}
{{\sqrt 2 (\nu ^2  - 1)}}\Psi _1 \exp \left( { - i\frac{{3t}}
{2}} \right)} \right\}\exp \left[ { - i\frac{{\varepsilon _1 ^2 }}
{{4(\nu ^2  - 1)}}t} \right] \hfill \\
\end{gathered} \label{eq72}
\]
\end{widetext}

\noindent where $\psi _0\;$and $\psi _1$ are the eigenfunctions of the unperturbed
Schr\"{o}dinger's operator $\left( {\xi (t)=\varepsilon _1\cos
t,\,0<\varepsilon _1<<1}\right) $. In this case
\begin{equation}
\begin{gathered}
\delta (t) =  - \frac{{i\varepsilon _1 }}
{{2\sqrt 2 }}\left( {\exp (it)\sin t + t} \right) \hfill \\
i\int\limits_0^t {\delta ^2 (z)\exp ( - 2iz)dz}  =  \hfill \\
\,\,\,\,\,\,\,\,\,\,\,\,\,\,\,\,\,\,\,\,\,\, - i\frac{{\varepsilon _1 ^2 }}
{8}\left[ {\frac{1}
{2}\left( {t - \frac{{\sin 2t}}
{2}} \right) + t^2 \exp ( - it)\sin t} \right] \hfill \\
\hfill \\
\end{gathered}
\end{equation}
\noindent so that within non-dimensional time interval $t\sim 1/\varepsilon
$ the expansion of the exact solution with accuracy up to the
first order of smallness has the form
\begin{eqnarray}
\Psi _0(x,t) &=&\pi ^{-{1/4}}\exp \left\{ {-{\dfrac{{it}}{{2}}}-{\dfrac{{x^2}%
}{{2}}}-}\right. \left( {{\dfrac{{\varepsilon _1t}}{{4}}}}\right) ^2+ \nonumber \\
&&\left. ix%
\dfrac{{\varepsilon _1t}}{{2}}\exp (-it)+
{\left( {{\frac{{\varepsilon _1t}}{{4}}}}\right) ^2\exp (-2it)}%
\right\} \,.  \label{eq75}
\end{eqnarray}

\section{Charged linear oscillator in an adiabatic homogeneous field}

In order to demonstrate application of the suggested theory, let us solve
the problem of motion of a particle in the field of a parabolic potential
subjected to a variable external force. Let $\varepsilon =\nu /\omega $ be a
small parameter $\left( {0<\varepsilon <<1}\right) $ in the adiabatic case.
Let us also take that parameter $\varepsilon _1$ is of order of unity. The
perturbation operator $\xi (\varepsilon t)x$ can not be taken to be small in
any way, cf. \cite{608}, that is, the problem of the perturbation theory can
not principally be solved by the methods of spectral analysis of operators.

To make calculations more transparent, it is worthwhile carrying out the
calculations from the very beginning rather than to use the resulting
formula (\ref{eq61}). For a fixed time instant $t$ problem (\ref{eq44}) is
written in the form
\begin{equation}
\left[ {-{\frac{{1}}{{2}}}{\frac{{d^2}}{{dx^2}}}+{\frac{{1}}{{2}}}x^2-\xi
(\varepsilon t)x}\right] \Psi _n(x,t)=E_n(t)\Psi _n(x,t)  \label{eq76}
\end{equation}
and has the following solution
\begin{eqnarray}
\Psi _n(x,t) &=(2^n\sqrt{\pi }n!)^{-{{1}/2}}\exp \left[ {-{{\left(
{x-\xi
(\varepsilon t)}\right) ^2}}/2}\right]\nonumber \\
&\times H_n\left( {x-\xi (\varepsilon t)}%
\right) ,  \nonumber \\
&\omega _n = n+1/2-\xi ^2/2\,,  \label{eq77}
\end{eqnarray}
where $H_n(z)$ denote Hermite polynomials \cite{613}, \cite{628}.

Using the recurrent relationships for Hermite polynomials it is easy to
calculate the matrix elements $v_{mn}(t)$ for $m\neq n$
\begin{equation}
v_{mn}(t)=-\left( {i}\dot{\xi}/2\right) \left[ {\sqrt{n+1}\delta _{m,n+1}-%
\sqrt{n}\delta _{m,n-1}}\right]  \label{eq78}
\end{equation}
where a dot denotes the total derivative with respect to time, so that $\dot{%
\xi}\sim \varepsilon $ and $\delta _{m,n}$ denotes Kronecker's delta. Since
the eigenfunctions (\ref{eq77}) are real-valued, the diagonal matrix
elements $v_{nn}=0,$ i.e. $\Omega _{n}=\omega _{n}=n+1/2-\xi ^{2}/2$.

With the help of eq. (\ref{eq78}) we can easily calculate values $\Delta
\Omega _{k}=\Delta \omega _{k}=-\dot{\xi}/2$ and coefficients $\bar{c}_{k}$,
see eq. (\ref{eq56})
\begin{equation}
\bar{c}_{k}=A_{k}\exp \left\{ {i{\int\limits_{{}}^{{}}{{\left[ \dot{\xi}{{%
^{2}(\varepsilon z)}/2}\right] }dz}}}\right\} \,.  \label{eq79}
\end{equation}
In the case under consideration $\langle H_{1}\rangle =0$ and $\tilde{H}%
_{1}=H_{1}$, then we find $\left\{ H_{1}\right\} $ and $\dfrac{{\partial {%
\left\{ {H_{1}}\right\} }}}{{\partial \bar{c}_{k}^{\ast }}}$%

\begin{eqnarray}
\left\{ {H_{1}}\right\} &=&-i{\sum\limits_{n,m}{^{\prime }}}\frac{{v_{mn}(t)}%
}{{\omega _{mn}(t)}}\bar{c}_{n}\bar{c}_{m}^{\ast }\exp \left[ {i\omega
_{mn}(\tau )t}\right] ,  \nonumber \\
\frac{{\partial {\left\{ {H_{1}}\right\} }}}{{\partial \bar{c}_{k}^{\ast }}}
&=&-{\sum\limits_{n}{^{\prime }}}\frac{{v_{kn}(t)}}{{\omega _{kn}(t)}}\bar{c}%
_{n}\exp \left[ {i\omega _{kn}(\tau )t}\right] \,.  \label{eq80}
\end{eqnarray}

Making use of relationships (\ref{eq80}) we obtain

\begin{multline}
c_{k}^{(2)}=A_{k}^{(2)}\exp \left[ {i{\int\limits_{0}^{t}{{\frac{\dot{\xi}%
^{2}}{{2}}}dz}}}\right] -   \\
{\sum\limits_{n}{^{\prime }}}\frac{{v_{kn}(t)}}{{%
\omega _{kn}(t)}}A_{n}\exp \left[ {i{\int\limits_{0}^{t}{{\frac{\dot{\xi}^{2}%
}{{2}}}dz}}+i\omega _{kn}t}\right] \,.  \label{eq81}
\end{multline}

The equations for determining the integration constants have the form

\begin{equation}
\left. {c_{k}^{(2)}}\right| _{t=0}=\delta _{k0}=A_{k}-{\sum\limits_{n}{%
^{\prime }}}\left( {{\frac{{v_{kn}(t)}}{{\omega _{kn}(t)}}}}\right)
_{t=0}A_{n}\,.  \label{eq82}
\end{equation}

Assuming, as above, that ${{\dot{\xi}}}(0)=v_{kn}(0)=0$, we obtain $%
A_{k}=\delta _{k0}$ and
\begin{equation}
c_{k}^{(2)}=\delta _{k0}\exp \left[ {i{\int\limits_{0}^{t}{{\frac{\dot{\xi}%
^{2}}{{2}}}dz}}}\right] -\frac{{v_{k0}}}{{\omega _{k0}}}\exp \left[ {i{%
\int\limits_{0}^{t}{{\frac{\dot{\xi}^{2}}{{2}}}dz}}+i\omega _{k0}t}\right]
\,.  \label{eq83}
\end{equation}
As follows from eq. (\ref{eq78}) $v_{k0}=-i\dot{\xi}/\sqrt{2}$, thus
\begin{equation}
c_{k}^{(2)}=\left[ {\delta _{k0}+i{\frac{\xi }{\sqrt{2}}}\delta _{k1}\exp
(it)}\right] \exp \left[ {i{\int\limits_{0}^{t}{{\frac{\dot{\xi}^{2}{(z)}}{{2%
}}}dz}}}\right] \,.  \label{eq84}
\end{equation}
Inserting these equalities into expansion (\ref{eq46}) yields the
post-adiabatic approximation

\begin{multline}
\Psi ^{(2)}(x,t)=\left[ {\Psi _{0}(x-\xi )+{\frac{{i{\dot{\xi}}}}{\sqrt{2}}}%
\Psi _{1}(x-\xi )}\right]   \\
\times \exp \left\{ {-{\frac{{it}}{{2}}}+i{%
\int\limits_{0}^{t}{{\frac{{\xi ^{2}}}{{2}}}dz}}+i{\int\limits_{0}^{t}{{%
\frac{\dot{\xi}^{2}}{{2}}}dz}}}\right\}  \label{eq85}
\end{multline}

which coincides with the expansion of the exact solution (\ref{eq69}) within
time interval $\Delta t\sim 1/\varepsilon $, approximation (\ref{eq85}) not
being applicable for increased times.

As follows from eq. (\ref{eq85}), the probability of the oscillator
excitation is equal to $\dot{\xi}^{2}/2$, i.e. it is zero at extreme points
of function $\xi (t)$. For a Gaussian distribution, $\xi (t)\sim \exp (-\tau
^{2})$ and the excitation probability has its only maximum at $\tau =1/\sqrt{%
2}$.

In the traditional Born-Fock approximation, \cite{613}, \cite{629}, the
expansion coefficient corresponding to eq. (\ref{eq81}) is given by
\begin{equation}
c_{kn}=\delta _{kn}+\int\limits_0^t\frac{{1}}{{\omega _{kn}(t^{\prime })}}%
\left( {{\frac{{\partial V}}{{\partial t}}}}\right) _{kn}\exp \left[ {i{%
\int\limits_0^t{\omega _{kn}(t^{\prime \prime })dt^{\prime \prime }}}}\right]
dt^{\prime },  \label{eq86}
\end{equation}
where $V(x,t)=-\xi (t)x.$ Carrying out simple manipulations we obtain
\begin{equation}
c_{k0}=\delta _{k0}-\frac{{\delta _{k1}}}{\sqrt{2}}\int\limits_0^t{{\dot{\xi}%
}}\exp (it^{\prime })dt^{\prime }\,.  \label{eq87}
\end{equation}
Comparing the latter equation with eq. (\ref{eq84}) indicates that the
factor $\exp \left[ {i{\int\limits_0^t{\left( {{{{\dot{\xi}}^2(z)}}/2}%
\right) dz}}}\right] $ is absent in eq. (\ref{eq87}). Equations (\ref{eq87})
and (\ref{eq84}) coincide at time instant $t\sim 1$, when a change in $\dot{%
\xi}\left( \varepsilon t\right) $ in the integrand can be neglected and
there are no conditions $\xi (0)={{\dot{\xi}}}(0)=0$ under which
approximation (\ref{eq84}) is valid.

\section{Adiabatic perturbation theory}

In the case of the
\index{adiabatic perturbation theory}adiabatic perturbation theory,
Schr\"{o}dinger's operator is as follows $%
\hat{H}=\hat{H}_{0}+\varepsilon \hat{V}(q,\tau )$ which allows us to carry
out calculations by using the results of Sec.6.6. However, as mentioned in
Sec. 6.2 it is more convenient to apply the formalism of the stationary
phase perturbation theory.

Indeed, by casting problem (\ref{eq62}) in terms of the eigenfunctions of
unperturbed Schr\"{o}dinger's operator $\hat{H}_{0}$ we find the effective
Hamilton function
\begin{equation}
\varepsilon H_{1}(c,c^{\ast },t,\tau )=-i\varepsilon {\sum\limits_{n,m}{%
^{\prime }}}v_{mn}(\tau )c_{n}c_{m}^{\ast }\exp \left( {i\Omega _{mn}^{0}t}%
\right) \,,  \label{eq88}
\end{equation}
where $v_{mn}(\tau )$ denote matrix elements of the perturbation operator
which are calculated by means of the unperturbed eigenfunctions, further $%
\Omega _{n}^{0}=\omega _{n}^{0}+\varepsilon \int\limits_{0}^{t}v_{nn}(\tau
)d\tau $ and $\Omega _{mn}^{0}=\Omega _{m}^{0}-\Omega _{n}^{0}$, $\omega
_{n}^{0}$ denote the eigenvalues of the unperturbed problem.

In this case formulae (\ref{eq52}) remain valid since the dependence of the
effective Hamilton function on slow time $\tau $ is observed only in terms
of the third order of smallness.

For problem (\ref{eq62}) we have

\begin{equation}
\begin{gathered}
\Psi _n^0 (x) = (2^n \sqrt \pi  n!)^{ - {1 \mathord{\left/
{\vphantom {1 2}} \right.
\kern-\nulldelimiterspace} 2}} \exp \left[ { - {{x^2 } \mathord{\left/
{\vphantom {{x^2 } 2}} \right.
\kern-\nulldelimiterspace} 2}} \right]H_n \left( x \right) \hfill \\
\omega _n^0  = n + {1 \mathord{\left/
{\vphantom {1 2}} \right.
\kern-\nulldelimiterspace} 2},\,\,\,\,\,\Omega _n^0  = \omega _n^0  = n + {1 \mathord{\left/
{\vphantom {1 2}} \right.
\kern-\nulldelimiterspace} 2} \hfill \\
v_{mn} (\tau ) =  - \frac{{\varepsilon _1 \xi (\tau )}}
{{\sqrt 2 }}\left[ {\sqrt {\frac{{n + 1}}
{2}} \delta _{m,n + 1}  + \sqrt {\frac{n}
{2}} \delta _{m,n - 1} } \right] \hfill \\
v_{nn} (\tau ) = 0 \hfill \\  \label{eq89}
\Delta \omega _k  =  - \frac{1}
{2}\varepsilon _1 ^2 \xi ^2 (\tau ) \hfill \\
\end{gathered}
\end{equation}

with parameter $\varepsilon _{1}$ being a small value of order of $%
\varepsilon $.

By virtue of the latter relationship in eq. (\ref{eq89}) we easily find the
second approximation to the expansion coefficients $c_n^{(2)}$ (under the
conditions $\xi (0)=\dot{\xi}(0)=0)$%
\begin{equation}
c_n^{(2)}=\exp \left[ {i\varepsilon _1^2{\int\limits_0^t{{\frac{{\xi ^2}}{{2}%
}}dz}}}\right] \delta _{k0}  \label{eq90}
\end{equation}
and the second approximation to the wave function
\begin{equation}
\Psi ^{(2)}=\psi _0(x)\exp \left[ {-{\frac{{it}}{{2}}}+i{\int\limits_0^t{{%
\frac{{\xi ^2}}{{2}}}dz}}}\right] \,,  \label{eq91}
\end{equation}
which coincides with adiabatic expansion (\ref{eq85}) when we take into
account a small factor at ${{\dot{\xi}}}$ due to parameter $\varepsilon _1$.

\section{Harmonic excitation of a charged oscillator. Non-resonant case}

To demonstrate the way of constructing solutions in the case of
non-stationary perturbation theory, we consider the case of a harmonic
external field which is frequently encountered in practical applications. In
this case function $\xi (t)$ in eq. (\ref{eq62}) takes the form $\xi
(t)=\varepsilon _1\sin \nu t$ and we can adopt that $\nu \neq 1$ and $\nu
\neq O(\varepsilon )$. In other words, we consider the non-resonant case and
parameter $\varepsilon _1$ is taken as having the order of smallness of
parameter $\varepsilon $.

Let us carry out the corresponding calculations in three ways: first,
stationary phase perturbation theory, second, non-stationary phase
perturbation theory developed here and finally, traditional method of \cite
{613}.

In the first case the matrix elements of the perturbation operator,
calculated by means of the unperturbed eigenfunctions, have the form
\begin{eqnarray}
v_{mn}(t) &=&\varepsilon _{1}(v_{mn}^{0}\exp (i\nu t)-v_{mn}^{0}\exp (-i\nu
t))\,,  \nonumber \\
v_{mn}^{0} &=&\frac{{i}}{{2\sqrt{2}}}\left[ {\sqrt{n+1}\delta _{m,n+1}+\sqrt{%
n}\delta _{m,n-1}}\right] \,.  \label{eq92}
\end{eqnarray}
The effective Hamilton function is given by
\begin{eqnarray}
\varepsilon _{1}H_{1}(c,c^{\ast },t) &=&-i\varepsilon _{1}{%
\sum\limits_{n,m}^{\infty }{^{\prime }}}v_{mn}^{0}\exp \left[ {i(\omega
_{mn}^{0}+\nu )t}\right] c_{n}c_{m}^{\ast }+  \nonumber \\
&&i{\sum\limits_{n,m}{^{\prime }}}v_{mn}^{0}\exp \left[ {i(\omega
_{mn}^{0}-\nu )t}\right] c_{n}c_{m}^{\ast }\,.  \label{eq93}
\end{eqnarray}
In this case $\bar{H}_{1}=\langle H_{1}\rangle =0$, so that $H_{1}=\tilde{H}%
_{1}$. The correction of the second order $\bar{H}_{2}$ to the averaged
Hamilton function is as follows
\begin{equation}
\bar{H}_{2}=-i\frac{{\varepsilon _{1}^{2}}}{{4(\nu ^{2}-1)}}\sum\limits_{n}%
\bar{c}_{n}\bar{c}_{n}^{\ast }\,.  \label{eq94}
\end{equation}
Next we find the second approximation for variables $c_{n}^{(2)}$%

\begin{multline}
c_{n}^{(2)}=\delta _{n0}-i\frac{{\varepsilon _{1}\nu }}{{\sqrt{2}(1-\nu ^{2})%
}}\;\delta _{n1}  \\
-i\frac{{\varepsilon _{1}}}{\sqrt{2}}\;\delta_{n1}\left[ {i\sin
\nu t-\nu \cos \nu t}\right] \exp (it)\,. \label{eq95}
\end{multline}

Taking into account these relationships we obtain the second approximation $%
\Psi ^{(2)}(x,t)$ to the solution of the non-stationary problem (\ref{eq62})
\begin{widetext}
\begin{eqnarray}
\Psi ^{(2)}(x,t) =\left\{ {{\left[ {\Psi _{0}+{\frac{{i\varepsilon _{1}}}{{%
\sqrt{2}(\nu ^{2}-1)}}}\Psi _{1}(i\sin \nu t-\nu \cos \nu
t)}\right] }\exp \left( {-{\frac{{it}}{{2}}}}\right) +}\right.
\left. {{\frac{{i\varepsilon \nu }}{{\sqrt{2}(\nu ^{2}-1)}}}\Psi
_{1}\exp
\left( {-i{\frac{{3t}}{{2}}}}\right) }\right\} \exp \left[ {-i{\frac{{%
\varepsilon _{1}^{2}}}{{4(\nu ^{2}-1)}}}t}\right] \,.  \label{eq96}
\end{eqnarray}
\end{widetext}
This result suggests that the spectrum remains equidistant and is
only
subjected to a common shift $\Delta \omega =\varepsilon _{1}^{2}/4\left( {%
\nu ^{2}-1}\right) $. In this case $\Delta \omega <0$ for $\nu <1$ ($\omega
<\omega _{0},$ $\omega $ being the frequency of excitation force), and $%
\Delta \omega >0$ for $\nu >1$ ($\omega >\omega _{0})$. The shift $\Delta
\omega $ at $\omega \rightarrow 0$ corresponds to effective elevation of the
bottom of the potential well. In the case of a high frequency external field
($\nu >>1)$ $\Delta \omega =\varepsilon _{1}^{2}/4{\nu ^{2}}$ that coincides
with the effective potential energy. Hence, the constructed expansion (\ref
{eq96}) coincides with the expansion of the exact solution (\ref{eq72}).

Let us now construct a solution by using formulae of the non-stationary
phase perturbation theory. In this case the matrix elements (\ref{eq48})
calculated by means of eigenfunctions (\ref{eq77}) of the instantaneous
Schr\"{o}dinger's operator have the form
\begin{equation}
v_{mn}(\tau )=-\frac{{i{\dot{\xi}}}}{\sqrt{2}}\left( {\sqrt{n+1}\delta
_{m,n+1}-\sqrt{n}\delta _{m,n-1}}\right)  \label{eq97}
\end{equation}
or
\begin{eqnarray}
v_{mn}(t) &=&v_{mn}^{0}\exp (i\nu t)+v_{mn}^{0}\exp (-i\nu t),  \nonumber \\
v_{mn}^{0} &=&-i\frac{{\varepsilon _{1}\nu }}{{2\sqrt{2}}}\left( {\sqrt{n+1}%
\delta _{m,n+1}-\sqrt{n}\delta _{m,n-1}}\right) \,.  \label{eq98}
\end{eqnarray}
As the eigenfunctions are real, then $v_{nn}=0$ and $\Omega _{n}=\omega
_{n}=n+\dfrac{{1}}{{2}}-\dfrac{{\xi ^{2}}}{{2}}$.

The effective Hamilton function is as follows
\begin{equation}
\begin{gathered}
\varepsilon _1 H_1 (c,c^ *  ,t) =  - i\varepsilon _1\sum\limits_{m,n = 0}^\infty  {'\{ v_{mn}^0 \exp \left[ {i(\omega _{mn}^0  + \nu )t} \right]  }  \hfill \\  \;\;\;\;\;\;\;\;\;\;\;\;\;\;\;\;\;\;\;\;\;\;\; + v_{mn}^0 \exp \left[ {i(\omega _{mn}^0  - \nu )t} \right]\} c_n c_m^ *\,.   \hfill
\end{gathered}
\end{equation}
It is evident that $\bar{H}_{1}=\langle H_{1}\rangle =0,$ so that $H_{1}=%
\tilde{H}_{1}$. The correction $\bar{H}_{2}$ to the averaged Hamilton
function, calculated by formulae (\ref{eq52}) with the help of eq. (\ref
{eq98}), is given by
\begin{equation}
\bar{H}_{2}=-i\frac{{\nu ^{2}}}{{4(\nu ^{2}-1)}}\sum\limits_{k}\bar{c}_{k}%
\bar{c}_{k}^{\ast }\,.  \label{eq100}
\end{equation}

The second approximation $\bar{c}_{k}$ for the evolutionary components of
variables $c_{k}$ is found from Hamilton's equations with Hamilton's
function $\bar{H}^{(2)}=\varepsilon _{1}\bar{H}_{1}+\varepsilon _{1}^{2}\bar{%
H}_{2}$ and has the form
\begin{equation}
\bar{c}_{k}=A_{k}\exp \left[ {-i{\frac{{\varepsilon _{1}^{2}\nu ^{2}}}{{%
4(\nu ^{2}-1)}}}t}\right] \,.  \label{eq101}
\end{equation}
Constants $A_{k}$ should be obtained from the initial conditions.

The second approximation $c_{k}^{(2)}$ to the original variables $c_{k}$ is
as follows
\begin{equation} \label{eq102}
\begin{gathered}
c_k^{(2)}  = \left\{ {A_k  - \varepsilon _1 \sum\limits_n {v_{kn}^0 A_n \left[ {\frac{{\exp \left[ {i\left( {\omega _{kn}  + \nu } \right)t} \right]}}
{{\omega _{kn}  + \nu }}} \right.} } \right. \hfill \\
\,\,\,\,\,\,\,\,\,\,\,\,\,\,\,\,\,\,\,\left. {\left. { + \frac{{\exp \left[ {i\left( {\omega _{kn}  - \nu } \right)t} \right]}}
{{\omega _{kn}  - \nu }}} \right]} \right\}\exp \left[ { - i\frac{{\varepsilon _1 ^2 \nu ^2 }}
{{4(\nu ^2  - 1)}}t} \right] \hfill \\
\end{gathered}
\end{equation}
For determining the integration constants $A_{k}$ from eq. (\ref{eq102}) we
obtain the following equation

\begin{equation} \label{eq103}
\begin{gathered}
\left. {c{}_k} \right|_{t = 0}  = \delta _{k0}  = A_k^{(2)}  - \varepsilon _1 \sum\limits_n {v_{kn}^0 \, \times \,}  \hfill \\
\,\,\,\,\,\,\,\,\,\,\,\,\,\,\,\,\,\,\,\,\,A_n^{(1)} \left[ {\left( {\omega _{kn}  + \nu } \right)^{ - 1}  + \left( {\omega _{kn}  - \nu } \right)^{ - 1} } \right] \hfill \end{gathered}
\end{equation}

\noindent where $A_{k}^{(1)}$ and $A_{k}^{(2)}$ are the first and the second
approximations to coefficients $A_{k}$ with respect to parameter $%
\varepsilon _{1}$. As the first approximation we can take $A_{k}^{(1)}={%
\delta _{k0}}$, then
\begin{equation}
A_{k}^{(2)}=\delta _{k0}-i\frac{{\varepsilon _{1}\nu }}{{\sqrt{2}(1-\nu ^{2})%
}}\delta _{k1}\,,  \label{eq104}
\end{equation}
where it is taken into account that $v_{k0}^{0}=-i\varepsilon _{1}\nu \delta
_{k1}/2\sqrt{2}$ and the final expression for $c_{k}^{(2)}$ takes the form
\begin{equation} \label{eq105}
\begin{gathered}
c_k^{(2)}  = \left\{ {\delta _{k0}  + i\frac{{\varepsilon _1 \nu }}
{{\sqrt 2 (\nu ^2  - 1)}}\delta _{k1}  + i\frac{{\varepsilon _1 \nu \exp (it)}}
{{\sqrt 2 (\nu ^2  - 1)}}} \right. \hfill \\
\left.\left. {\,\,\,\,\,\,\,\,\,\,\,\,\,\,\,\, \times \delta _{k1} \left[ {\cos \nu t - i\nu \sin \nu t} \right]} \right\}\exp \left[ { - i\frac{{\varepsilon _1 ^2 \nu ^2 }}
{{4(\nu ^2  - 1)}}t} \right] \right.\hfill \\
\end{gathered}
\end{equation}
Then we find the expression for phases in expansion (\ref{eq46})
\begin{equation}
-i\int\limits_{0}^{t}\omega _{n}(t)dt=i\frac{{\varepsilon ^{2}}}{{4}}t-i(n+%
\frac{{1}}{{2}})t+O(\varepsilon ^{2}).  \label{eq106}
\end{equation}
Inserting eqs. (\ref{eq105}) and (\ref{eq106}) into expansion (\ref{eq46})
we finally obtain
\begin{widetext}
\begin{equation} \label{eq107}
\begin{gathered}
  \Psi ^{(2)} (x,t) = \left\{ {\left[ {\Psi _0  + \frac{{i\varepsilon _1 }}
{{\sqrt 2 (\nu ^2  - 1)}}\Psi _1 (i\sin \nu t - \nu \cos \nu t)} \right]} \right. \hfill \\
\,\,\,\,\,\,\,\,\,\,\,\,\,\,\,\,\,\,\,\,\,\,\,\,\,\,\,\,\,\,\,\,\,\,\,\,\,
\,\,\,\,\,\,\,\,\,\,\,\,\,\,\,\,\,\,\,\,\,\,\,\,\,\,\,\,\,\,\,\,
\,\,\,\,\,\,\,\,\,\,\,\,\,\,\times \,\,\exp \left( { - \frac{{it}}
{2}} \right) + \left. {\frac{{i\varepsilon _1 \nu }}
{{\sqrt 2 (\nu ^2  - 1)}}\Psi _1 \exp \left( { - i\frac{{3t}}
{2}} \right)} \right\}\exp \left[ { - i\frac{{\varepsilon _1 ^2 }}
{{4(\nu ^2  - 1)}}t} \right] \hfill \\
\end{gathered}
\end{equation}
\end{widetext}

\noindent coinciding with eqs. (\ref{eq96}) and (\ref{eq72}).

Finally, taking into account that the second approximation of the present
analysis coincides with the first approximation of the traditional approach,
we obtain, for the case ${a_{k}=c_{k}}$, by means of the standard formulae
of the non-stationary perturbation theory, that
\begin{multline}
a_{k}=a_{k}^{(0)}+a_{k}^{(1)}= \\  \delta
_{k0}+i\frac{{\varepsilon _{1}\delta _{k1}}}{{\sqrt{2}(\nu
^{2}-1)}}\left( {i\sin \nu t-\nu \cos \nu t}\right) \exp (it)\,.
\label{eq108}
\end{multline}
Comparison of eqs. (\ref{eq105}) and (\ref{eq108}) allows us to
indicate a
number of inaccuracies in the standard courses. Firstly, a phase multiplier $%
\exp [-i\varepsilon _{1}^{2}\nu ^{2}t/$ $4{(\nu ^{2}-1)}],$ which is of
crucial importance for investigation of the coherent processes, is absent in
expression (\ref{eq108}). It can be neglected within a non-dimensional time
interval $t\sim 1$, but not within asymptotical intervals $t\sim
1/\varepsilon $. Thus, approximation (\ref{eq108}) and in turn the whole
solution is valid only within this small time interval.

Secondly, only the first approximation $a_k^{(1)}=\delta _{k0}$ rather than $%
a_k$ is subject to the initial condition (\ref{eq103}). This explains the
absence of the term $i\varepsilon _1\nu /{\sqrt{2}(\nu ^2-1)}$. Indeed, by
assuming $a_k^{(0)}=\delta _{k0}+\varepsilon _1\tilde{a}_{k0}^{(0)}$ and
subjecting the whole coefficient $a_k$ to the initial condition $\left. {a_k}%
\right| _{t=0}=\delta _{k0}$ we find
\begin{equation}
\tilde{a}_{k0}^{(0)}=i\frac{{\nu }}{{\sqrt{2}(\nu ^2-1)}}\delta _{k1}\,.
\label{eq109}
\end{equation}
Finally, as follows from solution (\ref{eq107}) or (\ref{eq72}) it is this
absent correction (rather than $a_k^{(1)}$) that determines the probability
of transition to the excited state.

\section{Harmonic excitation of an oscillator. Transition through a resonance%
}

Let us consider excitation of an oscillator by a weak resonant harmonic
field $V\left( x,t\right) =-{\varepsilon _{1}x\cos \nu t}$. In this case $%
1-\nu =\varepsilon $ $\left( 0\ll \varepsilon <1\right) $ and parameter ${%
\varepsilon _{1}}$ is a small value.

The effective Hamilton function constructed by means of the eigenfunctions
of the unperturbed Schr\"{o}dinger's operator has the form

\begin{widetext}
\begin{multline}
\varepsilon H_1(c,c^{*},t)=i\frac{{\varepsilon _1}}{{2\sqrt{2}}}%
\sum\limits_n\left\{ {\sqrt{n}c_nc_{n-1}^{*}\exp {\left[ {-i\varepsilon t}%
\right] }+\sqrt{n+1}c_nc_{n+1}^{*}\exp {\left[ {i\varepsilon t}\right] }}%
\right\}
\\
+ i\frac{{\varepsilon _1}}{{2\sqrt{2}}}\sum\limits_n\left\{ {\sqrt{n}%
c_nc_{n-1}^{*}\exp {\left[ {-i\left( {\nu +1}\right) t}\right] }+\sqrt{n+1}%
c_nc_{n+1}^{*}\exp {\left[ {i\left( {\nu +1}\right) t}\right] }}\right\} \,.
\label{eq110}
\end{multline}
\end{widetext}

In the first approximation we obtain the averaged Hamilton
function
\begin{equation} \label{eq111}
\begin{gathered}
  \varepsilon \bar H_1  = i\frac{{\varepsilon _1 }}
{{2\sqrt 2 }}\sum\limits_n {\left\{ {\sqrt n \,\bar c_n \,\bar c_{n - 1}^ *  \exp \left[ { - i\varepsilon t} \right]} \right.}  \hfill \\
  \left. {\,\,\,\,\,\,\,\,\,\,\,\,\,\,\,\,\,\,\,\,\,\,\,\,\,\,\,\,\,\,\,\,\,\,\,\,\,\,\,\,\,\,\,\,\,\,\,\, + \,\sqrt {n + 1} \,\bar c_n \,\bar c_{n + 1}^ *  \exp \left[ {i\varepsilon t} \right]} \right\} \hfill \\
\end{gathered}
\end{equation}
and the equation of first approximation for the evolutionary components $%
\bar{c}_k$ of the original variables $c_k$ takes the form
\begin{equation}
\stackrel{\cdot }{\bar{c}_k}=i\frac{{\varepsilon _1}}{{2\sqrt{2}}}\left[ {%
\sqrt{k+1}\bar{c}_{k+1}\exp \left( {-i\varepsilon t}\right) +\sqrt{k}\bar{c}%
_{k-1}\exp \left( {i\varepsilon t}\right) }\right] \,.  \label{eq112}
\end{equation}

It is easy to prove by direct differentiation that the solution of
the equation
\[
\dot{c}_{k}=i\frac{{\xi (t)}}{\sqrt{2}}\left[ {\sqrt{k+1}c_{k+1}\exp \left( {%
-i\varepsilon t}\right) +\sqrt{k}\exp \left( {i\varepsilon t}\right) c_{k-1}}%
\right]
\]
has the following form

\begin{equation*}
c_{0}(t) = \exp \left\{ {i\varepsilon
{\int\limits_{0}^{t}{dz\delta ^{2}(z)\exp (-2i\varepsilon
z)}}+{\frac{{1}}{{2}}}\delta ^{2}(t)\exp (-2i\varepsilon
t)}\right\} \,,
\end{equation*}
\begin{eqnarray}
c_{k}(t) &=&\frac{{{\left[ {-\delta (t)}\right] }^{k}}}{{\sqrt{k}!}}%
c_{0}(t)\,,  \nonumber \\
\delta (t)&=&-\frac{{i}}{\sqrt{2}}\int\limits_{0}^{t}\xi (z)\exp
(i\varepsilon z)dz\,.  \label{eq113}
\end{eqnarray}

Using this solution one can investigate the transition of the system through
the resonance.

We restrict our further investigation to the case of exact resonance $\left(
{\varepsilon =0}\right) $. Using relationship (\ref{eq113}) we find
coefficients $\bar{c}_{k}$%
\[
\bar{c}_{k}(t)=\frac{{1}}{\sqrt{k!}}\left( {{\frac{{i\varepsilon _{1}t}}{{2%
\sqrt{2}}}}}\right) ^{k}\exp \left[ {-\left( {{\frac{{\varepsilon _{1}t}}{{4}%
}}}\right) ^{2}}\right]
\]
and solution $\Psi (x,t)$ in the first approximation
\begin{eqnarray}
\Psi (x,t) &=&\sum\limits_{k=0}^{\infty }\frac{{1}}{\sqrt{k!}}\left( {{\frac{%
{i\varepsilon _{1}t}}{{2\sqrt{2}}}}}\right) ^{k}\exp \left( {-{\frac{{%
\varepsilon _{1}t}}{{4}}}}\right) ^{2}(2^{k}\sqrt{\pi }k!)^{-{1/2}}\times
\nonumber \\
&&\exp \left( {-{\frac{{x^{2}}}{{2}}}}\right) H_{k}(x)\exp \left[ {-i\left( {%
k+{\frac{{1}}{{2}}}}\right) t}\right] \,.  \label{eq114}
\end{eqnarray}
Carrying out summation in eq. (\ref{eq114}) by means of the generating
function for Hermite polynomials \cite{628}
\begin{equation}
\exp \left( {2xz-z^{2}}\right) =\sum\limits_{k=0}^{\infty }\frac{{z^{k}}}{{k!%
}}H_{k}(x),  \label{eq115}
\end{equation}
we obtain the final solution for the case of exact resonance
\begin{equation}
\begin{gathered}
\Psi (x,t) = \pi ^{ - \frac{1}
{4}} \exp \left\{ { - \frac{{it}}
{2} - \frac{{x^2 }}
{2} - \left( {\frac{{\varepsilon _1 t}}
{4}} \right)^2 } \right. \hfill \\
\left. {\,\,\,\,\,\,\,\,\,\,\,\,\,\,\,\,\,\,\, + ix\frac{{\varepsilon _1 t}}
{2}\exp ( - it) + \left( {\frac{{\varepsilon _1 t}}
{4}} \right)^2 \exp ( - 2it)} \right\} \hfill \\
\end{gathered} \label{eq116}
\end{equation}
This solution coincides with the expansion of the exact solution (\ref{eq75}).
    The probability of excitation of the oscillator has the form of a
Poisson distribution

\begin{equation}
w_{n}(t)=\left| {c_{n}(t)}\right| ^{2}=\frac{{\left( {\bar{n}}\right) ^{n}}}{%
{n!}}\exp (-\bar{n})\,,  \label{eq117}
\end{equation}

where $\bar{n}=\left( {{{\varepsilon _{1}t/2}}}\sqrt{2}\right) ^{2}$.

\begin{acknowledgments}
I wish to acknowledge the support of the author from
Prof. B.Matisov.
\end{acknowledgments}

\end{document}